\documentclass[11pt,doublespacing]{article}

\RequirePackage[OT1]{fontenc}

\RequirePackage[aos,amsthm,amsmath,natbib]{imsart}

\startlocaldefs
\theoremstyle{plain}
\newtheorem{theorem}{Theorem}[section]

\theoremstyle{plain}
\newtheorem{definition}{Definition}[section]

\theoremstyle{plain}
\newtheorem{example}{Example}[section]

\theoremstyle{plain}
\newtheorem{lemma}{Lemma}[section]

\theoremstyle{plain}
\newtheorem{corollary}{Corollary}[section]

\usepackage{amssymb}
\usepackage{graphics,epsfig,epic,ecltree}

 \newcommand{\indep}{ \_\hskip-4pt \sqcup \hskip-5pt\_}

\endlocaldefs

\begin{document}

\begin{frontmatter}
  \title{The Causal Manipulation of Chain
    Event Graphs} \runtitle{Chain event graphs} \author{\fnms{Eva}
    \snm{Riccomagno}\ead[label=e1]{riccomagno@dima.unige.it
    }} 
  \address{Via Dodecaneso, 35 \\ Genova 16149, Italy\\
    printead{e1}} \affiliation{Department of Mathematics, Universit\`a degli Studi di Genova} \and \author{\fnms{Jim Q.}
    \snm{Smith}\ead[label=e2]{j.q.smith@warwick.ac.uk}
  } 
  \address{ Gibbet Hill Road \\
    Coventry CV4 7AL, UK\\ printead{e1}} \affiliation{Department of
    Statistics, The University of Warwick}
  \runauthor{Riccomagno and Smith}

\begin{abstract}
  Discrete Bayesian Networks (BN's) have been very successful as a
  framework both for inference and for expressing certain causal
  hypotheses. In this paper we present a class of graphical models
  called the chain event graph (CEG) models, that generalises the
  class of discrete BN models. It provides a flexible
  and expressive framework for representing and analysing the
  implications of causal hypotheses, expressed in terms of the effects
  of a manipulation of the generating underlying system. We prove
  that, as for a BN, identifiability analyses of causal effects can be
  performed through examining the topology of the CEG graph, leading
  to theorems analogous to the back-door theorem for the BN.
\end{abstract}


\end{frontmatter}
\section{Introduction}

Bayesian networks have now been extended to Causal Bayesian Networks
(CBN's) using a non-parametric representation based on structural
equation models
\cite{GlymourCooper99,Pearl95,Pearl2000,Spirtesetal93}.  These
provide a framework for expressing assertions about what might
happen when the system under study is externally manipulated and
some of its variables are assigned certain values. Motivated by
comments in \cite{Shafer}, we develop an alternative graphical
representation of a causal model, called \emph{chain event graph
model}. This is constructed from an event tree together with a set
of ex-changeability assumptions. It can be seen as a generalisation
of a probability graph \cite{Bryant,Shafer} and typically has many
less nodes than the original event tree. It was introduced in
\cite{SmithAndersonCEG} in parallel with the present paper.
In \cite{SmithAndersonCEG} analogues to d-separation theorems that give
sufficient conditions for determining whether a conditional
independence statement holds, are given for CEG. Here
a causal extension of CEG models is discussed
which is as transparent and compelling as the extension from BN to CBN is.

We refer to the introduction in \cite{SmithAndersonCEG} for a
comparison of CEG models with other frameworks for propagations and
for interrogating an elicited model like probability decision graphs
\cite{Freidman,Jaeger}, context-specific networks \cite{Boutilier},
cofactors \cite{PZ} and case-factor diagram
\cite{McAllesteretal2004}. Here briefly we outline some of the reasons why
chain event graph models are important.

In some applications e.g. in Bayesian decision analysis
\cite{French89}, risk analysis \cite{BedfordCooke01}, physics
\cite{Lyons90}, biological regulation \cite{Churchill95}, often the
first stage of the elicitation of a model is based on the
elicitation of an event tree. Here is an example of the type of context we have in mind.

\begin{example} \label{Brick}\normalfont{
The police hold a suspect $S$ who they believe threw a brick through the kitchen window and stole a large quantity of money. 
The police hope to bring $S$ to court (indicator $X_{1}$) but might, for certain legal technicalities, be forced to release him. 
It is uncertain that the suspect was at the scene when the money was stolen (indicator $X_{2}$), that he was the individual who threw the brick and stole the money (indicator $X_{3})$, that the forensic service will find glass matching the window glass on the clothing of $S$ (indicator $X_{4}$), that a witness will identify $S$ (indicator $X_{5}$) and whether $S$ will be convicted $C$ or released $R$ (the ``effect'' indicator of interest $X_{6}$). 
Unless the suspect is identified by the witness as the one who threw the brick, $S$ will not be convicted. 
The glass match is believed only to depend on whether $S$ was present at the crime scene or not whilst the quality of the witness identification is believed to depend on whether or not $S$ was as at the scene of the crime as well as on whether or not he threw the brick. The police will later learn whether they have to release $S$ before trial, whether the witness will identify $S$ and the results of the forensic test.

Such a problem has an event tree representation. However this is rather cumbersome. Recently it has been more usual to represent this problem using a BN. Thus given $S$ is not released (i.e. conditional on $X_{1}=1$) the following BN is consistent with the possible unfoldings of events as described above 
 \begin{equation} \label{BNbrick} \begin{array}{ccccc} X_{2} & \rightarrow & X_{3} & \rightarrow & X_{5} \\ & \searrow & \downarrow & & \downarrow \\ & & X_{4} & \rightarrow & X_{6} \end{array} \end{equation}  
 }\end{example}
 
\begin{example} [Continuation of Example~\ref{Brick}] \label{BrickBN} \normalfont{ Although the graphical representation in (\ref{BNbrick}) is illuminating, it is not ideal. 
First, it is partial because the sample space that includes $X_{1}$ is not naturally a product space. 
Thus if $S$ is released, forensic evidence will not be collected, and the witness will not be allowed to testify, so in this sense these variables do not exist under this contingency. Of course we can formally define $X_{4}$ and $X_{5}$ conditional on $X_{1}=0$ so that a BN is consistent with this story. One such candidate is given  
\begin{equation*} \begin{array}{ccccccc} 
X_{2} & \rightarrow & X_{3} & \rightarrow & X_{5} \\ 
& \searrow & \downarrow & & \downarrow \\ 
& & X_{4} & \rightarrow & X_{6} & \leftarrow & X_1 
\end{array} \end{equation*}
This stores well probabilities. However, as a qualitative representation of the the possible unfoldings of events---perhaps used as the basis for embodying causal conjectures---it is not ideal. See also Example~\ref{univexample} below.

But suppose we use this BN representation of the problem. Second, we note that it only conveys certain aspects of the story. For example the fact that $S$ could only throw the brick $\{X_{3}=1\}$ if he were present at the scene of the crime $\{X_{2}=1\}$ and that conviction $\left\{ X_{6}=1\right\}$ requires $\left\{ X_{5}=1\right\}$ are not expressed in the diagram. Moreover we might like to incorporate into the representation context specific information because it is informative about various causal hypotheses (see e.g. \cite{Dawidetal2002}). This is particularly so of models of biological regulatory mechanisms which typically contain many noisy ``and'' and ``or'' gates \cite{AndersonSmith}. Third, we might well be interested in the causal effect of, for example, forcing the witness to identify $S$ as the culprit $\left\{ X_{5}=1\right\}$ if a match $\left\{ X_{4}=1\right\}$ in the glass is found. This is not represented in the usual semantics of the BN above. Furthermore it will not necessarily be the case that the manipulations we want to represent correspond to setting the original variables in the idle system to certain values.
} \end{example}

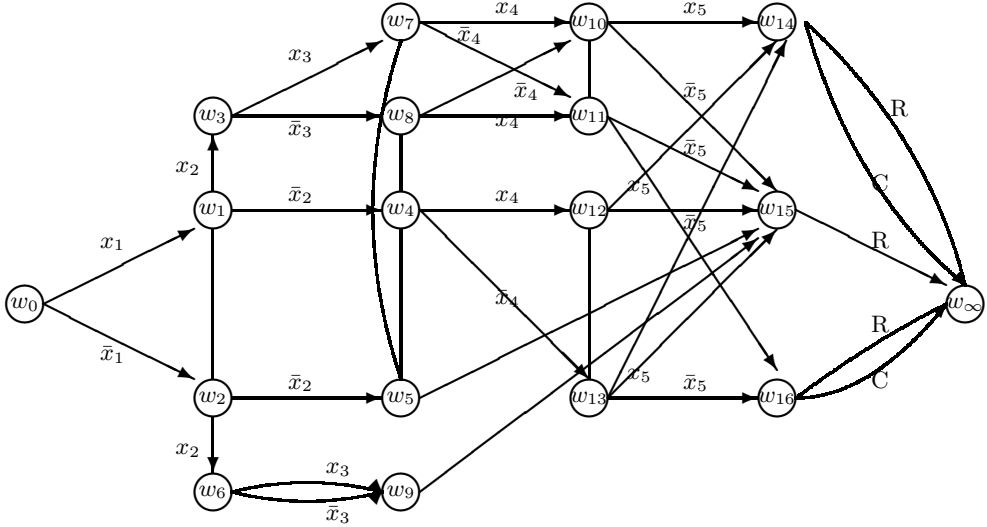
\begin{figure}
\setlength{\unitlength}{0.25cm}
\begin{picture}(50,32)(2,-14)\thicklines

\put(0,0){\circle{2}} \put(0,0){\makebox(0,0){$w_{0}$}}

\put(10,5){\circle{2}} \put(10,5){\makebox(0,0){$w_{1}$}}
\put(10,10){\circle{2}} \put(10,10){\makebox(0,0){$w_{3}$}}
\put(10,-5){\circle{2}} \put(10,-5){\makebox(0,0){$w_{2}$}}
\put(10,-10){\circle{2}} \put(10,-10){\makebox(0,0){$w_{6}$}}

\put(20,15){\circle{2}} \put(20,15){\makebox(0,0){$w_{7}$}}
\put(20,10){\circle{2}} \put(20,10){\makebox(0,0){$w_{8}$}}
\put(20,5){\circle{2}} \put(20,5){\makebox(0,0){$w_{4}$}}
\put(20,-5){\circle{2}} \put(20,-5){\makebox(0,0){$w_{5}$}}
\put(20,-10){\circle{2}} \put(20,-10){\makebox(0,0){$w_{9}$}}

\put(30,15){\circle{2}} \put(30,15){\makebox(0,0){$w_{10}$}}
\put(30,10){\circle{2}} \put(30,10){\makebox(0,0){$w_{11}$}}
\put(30,5){\circle{2}} \put(30,5){\makebox(0,0){$w_{12}$}}
\put(30,-5){\circle{2}} \put(30,-5){\makebox(0,0){$w_{13}$}}

\put(40,15){\circle{2}} \put(40,15){\makebox(0,0){$w_{14}$}}
\put(40,5){\circle{2}} \put(40,5){\makebox(0,0){$w_{15}$}}
\put(40,-5){\circle{2}} \put(40,-5){\makebox(0,0){$w_{16}$}}

\put(50,0){\circle{2}} \put(50,0){\makebox(0,0){$w_\infty$}}

\put(1,0){\vector(2,1){8}}   \put(4,3){$x_1$}
\put(1,0){\vector(2,-1){8}}  \put(4,-3){$\bar x_1$}

\put(10,6){\vector(0,1){3}}   \put(8,7){$x_2$}
\put(10,-6){\vector(0,-1){3}} \put(8,-8){$x_2$}
\put(10,4){\line(0,-1){8}}  

\put(11,10){\vector(2,1){8}}  \put(14,13){$x_3$}
\put(11,10){\vector(1,0){8}} \put(14,9){$\bar x_3$}
\put(11,5){\vector(1,0){8}}   \put(14,5.5){$\bar x_2$}
\put(11,-5){\vector(1,0){8}}  \put(14,-4.5){$\bar x_2$}
\qbezier(11,-10)(15,-9)(19,-10)  \put(19,-10){\vector(1,-1){0.1}}  \put(16,-9){$x_3$}
\qbezier(11,-10)(15,-11)(19,-10) \put(19,-10){\vector(1,1){0.1}}   \put(16,-11.5){$\bar x_3$}

\put(20,6){\line(0,1){3}}  
\put(20,-4){\line(0,1){8}}  
\qbezier(20,-4)(17,5)(20,14)

\put(21,15){\vector(1,0){8}} \put(25,15.5){$x_4$}
\put(21,15){\vector(2,-1){8}} \put(23,14){$\bar x_4$}
\put(21,10){\vector(1,0){8}} \put(26,11){$\bar x_4$}
\put(21,10){\vector(2,1){8}} \put(25,9.5){$x_4$} 
\put(21,5){\vector(1,0){8}} \put(25,5.5){$x_4$}
\put(21,5){\vector(1,-1){9}} \put(25,0){$\bar x_4$}
\put(21,-5){\vector(2,1){18}} 
\put(21,-10){\vector(4,3){18}} 

\put(30,-4){\line(0,1){8}}  
\put(30,11){\line(0,1){3}}  

\put(31,15){\vector(1,0){8}} \put(35,15.5){$x_5$}
\put(31,15){\vector(1,-1){9}} \put(35,11){$\bar x_5$}
\put(31,10){\vector(2,-1){8}} \put(35,8){$\bar x_5$}
\put(31,5){\vector(1,0){8}} \put(35,4){$\bar x_5$} 
\put(31,-5){\vector(1,0){8}} \put(35,-4.5){$\bar x_5$}
\put(31,5){\vector(1,1){9}} \put(32,6){$x_5$} 
\put(31,-5){\vector(1,2){9.5}} \put(32,-4){$x_5$}
\put(31,-5){\vector(1,1){9}} 
\put(31,10){\vector(2,-3){9}} 

\put(41,5){\vector(2,-1){8}} \put(45,3){R}
\qbezier(41.5,15)(48,9)(50,1) \put(50,1){\vector(1,-1){0.1}}  \put(46,10){R}
\qbezier(41.5,15)(44,6)(50,1) \put(50,1){\vector(1,-1){0.1}}  \put(45,6){C}
\qbezier(41,-5)(45,-2)(49,0) \put(49,0){\vector(1,1){0.1}}  \put(45,-1.5){R}
\qbezier(41,-5)(45,-5)(49,0) \put(49,0){\vector(1,1){0.1}}  \put(45,-4.5){C}

\end{picture}
\caption{CEG for Example~\ref{Brick}}
\label{BrickCEGfigure}
\end{figure}

The deficiencies of BN's as expressed in Example~\ref{BrickBN} should not be overemphasised. As stated in \cite{Dawidetal2002} there is an art to drawing the appropriate BN of a problem and it is sometimes necessary to redefine the variables defining the problem or add more edges on the graph to aid representation. For example were we to add an edge between $X_{4}$ and $X_{5}$ then the manipulation described in Example~\ref{BrickBN} can be expressed as a contingent decision (although we then have lost some information in the representation). Furthermore we can often transform the variables in a BN not only to encode more information but also so that the manipulation can be seen as setting this random variable to a value. Nevertheless the BN, whilst being consistent with a story like the one above, will still only be a partial representation in general. 
The CEG in Figure~\ref{BrickCEGfigure} gives a more expressive graphical representation, able to tell more of the story, which like the BN supports entirely graphical inferences about irrelevances and the potential effects of certain causal manipulations, albeit at the cost of some simplicity. 

The partially ordered sequence in which events unfold as expressed by the event tree
is retained in the CEG construction. Event trees explicitly
acknowledge asymmetries embedded in a structure both in its
development and in its sample space. These are retained in
the CEG. Recall that the dimension of the sample space often is critical for determining
identifiability especially when there are hidden variables, even in
symmetric models
\cite{SettimiSmith1998,SettimiSmith2000,SmithCroft}.

In probability trees conditional independence relations can be embedded through equations linking probabilities labelling the edges of
the tree. These are used to construct the vertices of the CEG and its undirected edges. Many conditional independence statements, including all those associated with BN's, many context specific BN's, and also over functions of the variables (such as noisy and/or gates), can be stated as equality of distributions associated with vertices on the underlying tree. This means that all such statements are expressed \emph{explicitly} via the topology of the CEG. See Example \ref{exindep} for complete independence models.

A basic assumption of our definition of causation (see Definition
\ref{manipulation}) is the belief that intervention or
manipulation of one or more vertices of the tree/CEG can model
external intervention on the underlying process being modelled.
This is analogous to the do-operator in \cite{Pearl2000} and is consistent with the representation of the effect 
of a cause on a tree model in \cite{Shafer}, albeit there not necessarily as a result of a manipulation.
It allows us to include naturally information relative to the
background idle system in the analysis of causal effects.

An advantage of the use of a framework based on event trees, and hence of CEG's,
is that causal hypotheses can be explicitly separated from any
direct link with the measurement process as will be illustrated later.

For a good discussion of many of the above points see \cite{Shafer},
in particular on the advantages of event trees (and hence CEG's) for coding asymmetrical
problems, and as powerful expression of an observer's beliefs especially
when those beliefs are based on an underlying conjecture about a
causal mechanism

Section~\ref{sectionCEGdefinition} contains the basic terminology
and definitions.  The extension from event trees and CEG's to causal
probability trees and causal CEG is in
Section~\ref{SectionSimulators}. A theorem of identifiability
analogous to the back-door theorem is proved in
Section~\ref{sectionBDtheorem}. Motivating and illustrative
examples are presented throughout.

\section{Chain event graphs}\label{sectionCEGdefinition}

From a Bayesian perspective probability trees describe the
observer's beliefs about what will happen as events unfold. The
edges out of a node $v$ of the probability tree represent the
possible unfolding that can occur from the situation labelled by
$v$, or equivalently the event space of a random variable that can
be indexed by $v$.  The sample space of the experiment at a node is
given by the branches of the probability tree at that point.

Through two equivalence relations on the nodes of the tree, we
construct a new model structure, called a chain event graph, which
includes Bayesian Networks and which provides a natural framework
for defining causality.

We start with some definitions to set up notation and formalise ideas.
Some of these definitions are slightly non-standard for reasons that
will become apparent later in the paper.  Section
\ref{probabilitytreesection} draws strongly from
\cite{Shafer,RiccomagnoSmith2007,SmithAndersonCEG} and we refer to those works for further
details.

\subsection{Probability trees} \label{probabilitytreesection}
The model structure upon which a
manipulation operation is defined in
Section~\ref{SectionSimulators} is a graph constructed
from a rooted, directed (event) tree $\mathcal
T=(V(\mathcal{T}),E(\mathcal{T}))$ where $V(\mathcal{T})$ and
$E(\mathcal{T})$ are the set of vertices or nodes and of edges,
respectively. Write $V=V(\mathcal{T})$ and $E=E(\mathcal{T})$ when there is no
ambiguity and assume $V$ and $E$ finite. The single root node is denoted by $v_0$. Between two vertices there is at most on edge. Each edge $e\in
E$ is directed with a parent node $v$ and a child node $v^\prime$. We
write $e=(v,v^\prime)$ and note that this edge can be identified with its child vertex $v^\prime$.

\begin{definition}  For $v\in V$ let
$\mathbb{X}(v)=\{ v^\prime \in V $ such that there exists 
  $e\in E$ such that $e=(v,v^\prime)\} $ and call $\mathbb{X}(v)$ the set of children of $v$.
\end{definition}

Thus $\mathbb{X}(v)$ is in one-to-one correspondence with the set of
edges out of $v$. If $\mathbb{X}(v)$ is the empty set, then $v$ is
called a \emph{leaf node}, otherwise it is called a \emph{situation}.
The set of situations, $S=S( \mathcal{T})\subset V(\mathcal T)$, will
have particular significance. Note that $\left\{\mathbb{X(}v):v\in
  S\right\}\cup \{v_{0}\}$ partitions $V$.

A \emph{path} between two vertices $v$ and $v^\prime$ is an
ordered sequence of nodes
$\lambda=\lambda(v,v^\prime)=(v_1,\ldots,v_{n[\lambda]+1})$ where
$v_1=v$, $v_{n[\lambda]+1}=v^\prime$ and $v_k$ is the child of $v_{k-1}$
and the parent of $v_{k+1}$ for $k=2,\ldots,n[\lambda]$.
This path can equivalently be identified with the ordered sequence of its edges
$\lambda=(e_1,\ldots,e_{n[\lambda]})$ where
$e_k=(v_k,v_{k+1})$ for $k=1,\ldots,n[\lambda]$. The number $n[\lambda]$ of
edges in called the \emph{length} of the
path. We write $v\in \lambda$ whenever $\lambda$ contains the vertex $v$.

\begin{definition}
  Let $\mathbb{X}(\mathcal T)=\mathbb{X}=\left\{\lambda (v_{0},v): v\in
    V\backslash S \right\}$ be the set of
  root-to-leaf paths. The elements of $\mathbb{X}$ are called
  {\normalfont atomic events} of $\mathcal T$.
\end{definition}

Clearly, $\mathbb{X}(\mathcal T)$ is in one-to-one correspondence with the leaves of $\mathcal T$. 
Paths determine a partial order on $V(\mathcal{T})$.
A probability tree could be drawn so that the situations along each root-to-leaf path correspond to one of the possible historical developments of the modelled problem.
In some cases this directionality is inherent to the problem to be modelled and expresses a conjecture about the ordering in which one situation follows another (see in particular \cite[Section 2.8]{Shafer}).

\subsection{Primitive probabilities}
Next we impose a probability structure on $\mathcal T$. Consider
$\mathbb X$ as sample space, the power set $2^{\mathbb
X}$, and a probability $\operatorname{P}$ on $(\mathbb X,2^\mathbb X)$. Thus
$\operatorname{P}(\lambda)$  is the probability of the path $\lambda\in \mathbb X$.
A \emph{probability tree} is a directed tree $\mathcal T=(V,E)$ such
that to each situation $v\in  S(\mathcal T)$ is associated a
discrete random variable $X(v)$ whose sample space is $\mathbb
X(v)$. The distribution of $X(v)$, $v\in  S(\mathcal T)$, determines
the \emph{primitive probabilities}
\[
\pi(e)=\pi(v^\prime|v)=\pi(v^\prime)=\operatorname{P}(X(v)=v^\prime) \text{ for }
v^\prime \in \mathbb X(v), \text{ for } e=(v,v^\prime)
\]
which are fundamental in the paper. Let $\Pi_a(\mathcal T)$, or
simply $\Pi_a$, denote the set of all primitive probabilities and
$\pi(e)$ is the label or colour of $e$.

The random variable $X(v)$ can be interpreted as a function on $\mathbb X$ which assigns zero probability to the set of paths not through $v$. Furthermore, for $\lambda\in \mathbb X$ the formula $X(v)(\lambda)=v^\prime$ states that $X(v)$ maps $\lambda$ into the path through $v$ and $v^\prime$, hence the notation $\pi(v^\prime|v)$. This gives an interpretation of $X(v)$ as a function from $\mathbb X$ to $\mathbb X(v)$. Example~\ref{simpledag} provides another
motivation for this notation.
For every situation $v\in \mathcal S(\mathcal T)$
the sum-to-one condition can be written as
\begin{equation} \label{sumtoone}
\sum_{v^\prime \in \mathbb X(v)} \operatorname{P}(X(v)=v^\prime) =
\sum_{v^\prime \in \mathbb X(v)} \pi(v^\prime|v) =  1 .
\end{equation}

Random variables on vertices along a path are required to
be mutually independent. It then follows that the probability $\operatorname{P}(\lambda)$ of the atomic event
$\lambda=(e_1,\ldots,e_{n[\lambda]}) \in \mathbb X$ is the
following product of primitive probabilities
\begin{equation} \label{atomicprimitiveprobabilitiesformula}
\prod_{j=1}^{n[\lambda]} \pi(e_j)=\operatorname{P}(\lambda) .
\end{equation}
Furthermore if $\pi (v^{\prime}|v)=0$ for some
$v,v^\prime$ then any atomic event including $v^\prime$ has zero
probability of occurring. In some circumstances the branch starting 
at $v^\prime$ could be deleted from the tree. 

Above we have defined probabilities of the atomic events.
Subsequently we defined some random variables over the probability
space $(\mathbb X,2^\mathbb X,\operatorname{P})$ and finally used these random
variables to define the primitive probabilities $\Pi_a$. In practice
when modelling a problem over a probability tree we would usually
work the other way round. While constructing the tree from root to
leaves the modeller has a (possibly not definite) idea of the
distribution of the random variables sitting on situations. Then
Equation~(\ref{atomicprimitiveprobabilitiesformula}) is used to
determine the probabilities of the atomic events. The $\operatorname{P}$'s and 
$\pi$'s provide equivalent  parametrisations of the probability tree
model under the assumption of strict positivity. We do not discuss
this further, but give a simple example to show an algebraic reason
by which we prefer to parametrise the probability space $(\mathbb
X,2^\mathbb X,\operatorname{P})$ using the primitive probabilities $\Pi_a$.

\begin{example} \label{simpletree} \normalfont{
For the tree in Figure~\ref{simpletreeFig} Equations~(\ref{atomicprimitiveprobabilitiesformula}) are in the left-hand-side (LHS) of the tableau below
\[ \begin{array}{lll|lll}
p_2=1-\pi_1 & & p_5=\pi_1(1-\pi_3-\pi_4) & \pi_1=1-p_2 & & \pi_4=\displaystyle\frac{p_4}{1-p_2}\\
p_4=\pi_1 \pi_4 & & p_7=\pi_1\pi_3(1-\pi_6) & \pi_3=\displaystyle\frac{1-p_2-p_4-p_5}{1-p_2} \\
p_6=\pi_1 \pi_3 \pi_6 && & \pi_6=\displaystyle\frac{p_6}{1-p_2-p_4-p_5}
\end{array} \]
where $p_i$ is the probability of the path starting at the root vertex and ending in the vertex $v_i$, $i\in\{1,\ldots,7\}$ and $\pi_i=\pi(v_i)$. The equations in the LHS of the tableau give the probability of an atomic event $\lambda$ as a polynomial in the primitive probabilities of degree equal to the length of the path $\lambda$. 

The RHS of the tableau gives the primitives in terms of the atomic event probabilities. These are ratios of polynomials of degree one in the paths probabilities and are defined if the denominators are not zero.

Note that $p_6+p_7=\pi_1\pi_3$ is the probability of the path $\{v_0,v_2,v_3\}$, and that
$\pi_4=\displaystyle\frac{p_4}{1-p_2}=\displaystyle\frac{\operatorname{P}(X(v_0)=v_1,X(v_1)=v_4)}{\operatorname{P}(X(v_0)=v_1)}=\operatorname{P}(X(v_1)=v_4|X(v_0)=v_1)$.
Hence the notation $\pi_4=\pi(v_4|v_1)$ is used to underline its interpretation as the probability of reaching $v_4$ having arrived in $v_1$. See also Shafer  \cite{Shafer}.
We will see later that this directional parametrization is a natural one to use when modelling causal hypothesis just as the directed parametrization of the BN naturally projects into the CBN.
}\end{example}

The primitive probabilities, which might be unknown or partially known,
satisfy some logical constraints which often are algebraic. Obvious ones are $0\leq \pi(e)\leq 1$ and the linear polynomials in Equations~(\ref{sumtoone}). Supplementary constraints might be inherent to the logic of the problem been modelled and others might be imposed by the modeller, e.g. in Figure~\ref{simpletreeFig} the modeller might know that $X(v_1)$ follows a Binomial distribution with $\operatorname{P}(X(v_1)=v_3)=s^2$ and $\operatorname{P}(X(v_1)=v_5)=(1-s)^2$ for some $s\in [0,1]$. The definition of one class of such constraints leads to the notion of chain event graphs below.

\begin{example}[Continuation of Example~\ref{simpletree}]\label{simpletreealgebra} 
\normalfont{Assume the primitive probabilities and the path probabilities are all 
unknown indeterminates and consider the polynomial ideal generated by
$ G=\{
p_2-(1-\pi_1 ),  p_5-\pi_1(1-\pi_3-\pi_4),
p_4-\pi_1 \pi_4,
p_7-\pi_1\pi_3(1-\pi_6), p_6-\pi_1 \pi_3 \pi_6  \} 
$.
This is the infinite set of polynomials  of the form $\sum_{g\in G}
s_g g$ where $s_g$ is any polynomial in $p_2,p_4,p_5,p_6,p_7,
\pi_1,\pi_3,\pi_4,\pi_6$. The elimination ideal of
the $\pi$ variables is generated by the  polynomial condition
$\{p_2+p_5+p_4+p_7+p_6-1\}$, obviously. 
The constraint $\pi_4=\pi_3\pi_6$ is considered by adjoining the polynomial $\pi_4-\pi_3\pi_6$ to $G$.
The elimination ideal contains, obviously, the polynomial $p_4=p_6$. 
The propagation of polynomial constraints on the $\pi$'s and $p$'s 
is imposed by adjoining the corresponding polynomials to $G$ and computing the relevant elimination ideal.
This makes CEG's analysis amenable of techniques from algebraic statistics \cite{Mono}. 
Note that the manipulations defined in Section~\ref{puremanipulation} are polynomial constraints setting some $\pi$'s equal to zero. 
} \end{example}

\subsection{Stages}
\begin{definition}\label{stages} Let $\mathcal T$ be a probability
tree and $\Pi_a(\mathcal T)$ an associated set of primitive
probabilities. Two situations $v_1,v_2\in S(\mathcal T)$ are said to
be {\normalfont stage-equivalent} if and only if $X(v_1)$ and
$X(v_2)$ have the same distribution.
\end{definition}
Definition~\ref{stages} requires that there exists a one-to-one map
$\mu: \mathbb{X}(v_1) \rightarrow \mathbb{X}(v_2)$ and that
$\pi(v|v_1)=\pi(\mu(v)|v_2)$ for all $v\in \mathbb{X}(v_1)$.
It determines an equivalence relation on $\mathcal{S(T)}$ whose
equivalence classes are called \emph{stages}. For each stage $u$ define
\[ \Pi(u)=\{\pi (v^{\prime }|v):v^{\prime }\in \mathbb{X}(v) {\text{
for some }} v {\text{ representative of }} u\} \] and $\Pi=\Pi
(\mathcal{T})=\bigcup_{u\in L(\mathcal{T})} \Pi(u)$. The set of
primitive probabilities $\Pi$ is no larger than the set of all
primitive probabilities $\Pi_a$ and clearly still sufficient for
determining the distribution of all random variables measurable with
respect to $(\mathbb X,2^\mathbb X)$. The pair $(\mathcal{T},\Pi
(\mathcal{T)})$ is called a \emph{probability tree model}.

\subsection{Examples}
\begin{example}\label{ABCexample}\normalfont{
Suppose that only three incidents $A,B,C$ can occur and the order of occurrence of $B$ and $C$ is
relevant and contingent on whether $A$ or not $A$ ($\bar A$) happens.
Their history unfolds according to the probability tree in Figure \ref{stages_indep}.
For example the path
$(v_0,v_1,v_3)$ represents whether after $A$ has occurred, $B$
occurs. The random variable $X(v_0)$ is the indicator function of
the event of the incident $A$ happening; $\pi(v_1)$ is the probability of this event and the
primitive $\pi(v_3)$ is the probability that $B$ occurs if $A$ has
occurred. }\end{example}

Henceforth we shall assume that the tree fully represents all
possible unfoldings of situations. The unfolding 
``if first $A$ and then $B$, then $C$'' is simply not part of our story. 
The probability of the incident of $A$ does not happen and then $B$ does, 
is given by
$\pi(v_2)\pi(v_6)+\pi(v_2)\pi(v_7)\pi(v_{9})$ being the sum of the
probabilities of the atomic events $\{v_0,v_2,v_6\}$ and
$\{v_0,v_2,v_7,v_9\}$.
Note that the path $\sigma $-algebra is not the $\sigma$-algebra generated by $\{A,B,C\}$ where these are thought of as events, since this
 $\sigma$-algebra cannot express the ordering of incidents.


Figure~\ref{stages_indep} could represent the following
situation. A woman with epilepsy happens not want to conceive ($A$) or
 wants to conceive ($\overline A$). She has three alternatives:
to take medicine $B$, to take medicine $C$ or to follow some other
health regime, $\overline{B\cup C}$. Medicine $C$ has less side
effects than $B$ (thus we shall impose $\pi(v_4)>>\pi(v_3)$) but it
might have an adverse effect on the formation of the nervous system
of the foetus if she becomes pregnant, thus $\pi(v_7)< \pi(v_4)$. If
she discovers pregnancy within the first three months from
conception then the medicine $B$ can be taken as a supplement to $C$
to reduce greatly this adverse effect. But unless she discovers
pregnancy she would never take $B$ after having taken $C$ because of
other health risks to herself linked to the combined use of $B$ and
$C$.

\begin{example}[Continuation of Example~\ref{ABCexample}]\label{contABCexample}\normalfont{
The modeller might want to assert that $\pi(v_3)=\pi(v_7)$ and $\pi(v_4)=\pi(v_6)$. This assigns $v_1$ and $v_2$ to the same stage. This assertion implies two things.
First if the woman is not taking either $B$ or $C$ then the probability of changing her health regime is the same whether or not she wants to become pregnant. Second if she wants to become pregnant then she will prefer to take $C$ as much as she is likely to take $B$ if she does not want to become pregnant.
}\end{example}

\begin{example} \label{simpledag}{\normalfont
Let $X$ and $Y$ be two binary random variables.
Let $X=1$ be the event that a person watches a violent movie on Saturday
night and $Y=1$ the event of that person getting into a fight on Saturday
night. In Figure~\ref{Figure1} $X(v_0)=X$, $X(v_1)=(Y|X=1)$ and
$X(v_2)=(Y|X=0)$. The path $(v_0,v_1,v_3)$ corresponds to having watched
a violent movie and ending up in a fight and the path $(v_0,v_2,v_5)$
to not having watched a violent movie and ending up in a fight. The three
primitive probabilities $\pi(v_1)$, $\pi(v_3)$ and $\pi(v_5)$ are sufficient 
to parametrise the model. 
}\end{example}

\begin{example}[Continuation of Example \ref{simpledag}]
    \label{simpledagCont}{\normalfont In Figure
    \ref{Figure1} $v_1$ and $v_2$ can be in the same stage in two
    ways: (1) $\pi(v_3)=\pi(v_5)$ and (2) $\pi(v_3)=\pi(v_6)$. In (1)
    we assume that $X$ and $Y$ are independent, i.e. watching a violent movie has no effect on the probability of subsequent violence;
    while in (2) we assume
    that $\operatorname{P}(Y=0|X=0)=\operatorname{P}(Y=1|X=1)$, i.e. the probability that violence occurs after seeing violence can be equated with the probability of non-violence occurring after not seeing violence---a type of conservation of violence law.
 If the values $X$ and $Y$ take
    were $-1$ and $1$ then this last equality would imply the
    independence of $X$ and $XY$.
}\end{example}

\subsection{Chain event graphs}

In this section we assume that the observer is able to express two
pieces of qualitative information: the topology of the probability
tree and its stages. These two sources of information can be fully
represented using a mixed graph called a chain event graph. Mixed
means that the graph has directed and undirected edges. Directed
edges are labelled with primitive probabilities while undirected
edges are not labelled. 

Let $(\mathcal T, \Pi)$ be a probability tree model.
For a situation $v\in S$ let $\mathcal T(v)$ be the sub-trees starting at $v$.
In \cite{Lauritzen96} $\mathcal{T}(v)$ is called the subgraph induced by $v$ and its ancestors.
Let $\Pi_v$ be the subset of $\Pi$ labelling edges in  $\mathcal{T}(v)$.
Then  $(\mathcal{T}(v),\Pi_v)$ is a probability tree model.

\begin{definition} \label{position}Two situations  $v$ and $v^\ast$ in the probability tree
$(\mathcal{T}, \Pi)$ are equivalent if and only if
  \begin{enumerate}
  \item $\mathcal{T}(v)$ and $\mathcal{T}(v^\ast)$ are isomorphic. That is,
there exists a map  $\mu$ from the sets of vertices in $\mathcal T(v)$ and $\mathcal T(v^\ast)$ such that
$(\mu(v_1),\mu(v_2))$ is an edge in $\mathcal T(v^\ast)$ if and only if $(v_1,v_2)$ is an edge in $\mathcal T(v)$, and
  \item for every $w$ situation in $\mathcal{T}(v)$, $w$ and
    $\mu(w)$ are in the same stage.
  \end{enumerate}

  The induced equivalence classes are called {\normalfont positions}
  and $K(\mathcal{T})$ is the set of positions.
\end{definition}
Item~2. in Definition~\ref{position} simply means that 
$\pi(v_2|v_1)=\pi(\mu(v_2)|\mu(v_1))$ for all possible
    $v_1,v_2\in \mathcal{T}(v)$. 
In Definition~\ref{position} we require that the sub-trees are
topologically isomorphic and that their edge probabilities match
according to the isomorphism, whether they are known fixed values,
unknown fixed values or indeterminates. Clearly the partition of
situations into positions is a refinement of the partition into
stages: if two situations are in the same position then they are in
the same stage. Two situations are in the same position when the
processes governing the stories unfolding from them are believed to
be governed by processes with the same distribution. Note that this
is a predictive, not a retrospective equivalence. Once a unit
reaches the situation $v$ or the situation $v^\ast$, all pairs of
possible unfoldings from $v$ and $v^\ast$ occur with the same
probabilities. It is in this respect that positions are natural
objects on which to describe a causal manipulation.

In broad terms, stages are used for estimation as they reduce the
number of parameters (primitive probabilities) and positions are
used to express conditional independence statements
\cite{SmithAndersonCEG}. Positions are used to form the vertices of
a new graph called the chain event graph. Its undirected edges join
positions at the same stage and its directed paths correspond to
root-to-leaf paths of the probability tree model.

\begin{definition} Let $(\mathcal{T},\Pi)$ be a
  probability tree model. Its chain event graph, $\mathcal{C(T)}$, is
  the mixed graph $\left(
V(\mathcal{C(T)}), E_d(\mathcal{C(T)}), E_u(\mathcal{C(T)}),
\Pi(\mathcal{C(T)})
\right) $ where
  \begin{enumerate}
  \item  $V(\mathcal{C(T)})=K(\mathcal{T})\cup \{w_\infty \}$ is the vertex set. The vertex $w_\infty$ is called the {\normalfont
      sink vertex}.
  \item  $E_d(\mathcal{C(T)})$ is a multi-set of directed edges and is
    partitioned into two sets, $E_1(\mathcal{C(T)})$ and
    $E_2(\mathcal{C(T))}$ constructed as follows. For each $w\in
    K(\mathcal{T})$ choose $v\in V(\mathcal{T})$ a representative of
    $w$. For each $(v,v^\prime)$ edge in $E(\mathcal{T})$
    \begin{enumerate}
    \item if $v^\prime$ is in position $w^\prime$ then add a directed
      edge from $w$ to $w^\prime$ to the multi-set
      $E_1(\mathcal{C(T)})$,

    \item if $v^\prime$ is a leaf node then add a directed edge from
      $w$ to $w_\infty$ to the multi-set $E_2(\mathcal{C(T))}$.

     \end{enumerate}
  \item $E_{u}(\mathcal{C(T)})$ is a set of undirected edges joining positions in the same stage, namely
\[ \begin{array}{rcr}
E_{u}(\mathcal{C(T)}) &=& \left\{
\{ w,w^\prime \}:\text{ with }w\neq w^\prime \text{ and }
 \right. 
\text{ there exist } v,v^\prime\in S({\mathcal T}), 
\\ && \quad \left. u\in
  L(\mathcal {T})  \text{ with } v,v^\prime \in u
\text{ and } v\in w,
  v^\prime\in w^\prime \right\} .
\end{array} \]
  \item $\Pi(\mathcal C(\mathcal T))=\Pi$ and if $e_1,e_2\in E_d(\mathcal{C(T)})$ and $\pi(e_1) =\pi(e_2)$ in the original tree, then $e_1$ and $e_2$ have the same label in the CEG.

\end{enumerate}
\end{definition}
When there is no ambiguity we write $(V_\mathcal C, E_d,E_u,\Pi)$
and $\mathcal C$ for the chain event graph. The CEG fully expresses 
the structure $\mathbb X$ of the sample space of a tree because
there are as many directed root-to-leaf paths in the original tree
as there are root-to-sink paths in its CEG. However often it has
many less edges.

If two vertices $v$ and $v^\ast$ of the original tree are in the
same position, then for each path $\lambda(v,v_M)$ in the sub-tree
${\mathcal{T}}(v)$ there exists a corresponding path
$\lambda^\ast(v^\ast,v_M^\ast)$ in ${\mathcal{T}}(v^\ast)$ along 
which the same evolutions occurs. 
This implies that $\operatorname{P}(\lambda)= \operatorname{P}(\lambda^\ast)$. 
In particular consider the root-to-leaf paths, given in terms of vertices, 
$\lambda(v_0,\ldots,v,\ldots,v_M)$ and
$\lambda^\ast(v_0,\ldots,v^\ast,\ldots,v^\ast_{M^\ast})$ where $v_M$
and $v^\ast_{M^\ast}$ are leaves in $\mathcal T$ and $v$, $v^\ast$
are in the same position. Then
\begin{equation} \label{probpaths} 
\operatorname{P}(\lambda)  = \operatorname{P}(\lambda(v_0,v)) \operatorname{P}(\lambda(v,v_M)) \text{ and }
\operatorname{P}(\lambda^\ast)  = \operatorname{P}(\lambda^\ast (v_0,v^\ast)) \operatorname{P}(\lambda(v,v_M))  .
\end{equation}
The same formula holds when the paths are considered in the CEG. In this case
$v_0$ is substituted by the root node and $v_M$
and $v^\ast_{M^\ast}$ become the sink node.

\begin{example}[Continuation of Example~\ref{simpledag}] \label{exindep}
{\normalfont Figure \ref{CEGFigure1} gives the CEG when $v_1$ and
$v_2$ are in the same stage. The values of the edge labels indicates
whether it is model (1) or (2). }
\end{example}

\begin{example}{\normalfont Figures \ref{fax.eps} and \ref{CEGfax.eps}
    give a tree and its CEG for the stage partition $\left\{ \{v_0\}, \{
      v_1, v_3, v_{13}, v_{17} \}, \{ v_2, v_7 \}, \{ v_5, v_{9} \},
      \{v_{19}\} \right\}$ and the position partition $\left\{ \{v_0\}
    \right.$, $\{ v_1,v_3\},$ $ \left.\{ v_5, v_{9}\}, \{v_2\},
      \{v_7\}, \{v_{13}\}, \{v_{17}\}, \{v_{19}\}, w_\infty \right\}$.
  }
\end{example}

\begin{example}[Continuation of Example~\ref{contABCexample}] \normalfont{
The positions are $\{v_0\},\{v_1\},\{v_2\}, \{v_7\}, w_\infty$ and the CEG is in
Figure~\ref{ABCceg}, where $\pi_i=\pi(v_i)$.
} \end{example}

\begin{example}[Bayesian network] {\normalfont In
\cite{SmithAndersonCEG} it is proved that any discrete Bayesian
network on the random variables $\left\{ X_1,\ldots,X_n
\right\}$ can be expressed by a CEG. Example~\ref{fromCEGtoBN} below shows how to retrieve the conditional independence statements of the BN from the topology of any of its CEG's. Like
context specific BNs \cite{Boutilier} but unlike the probability
decision graph \cite{Jaeger} or the probability graph \cite{Bryant},
the CEG provides a generalisation of the BN. In particular
 two situations $v_{i-1}$ and $v^\prime_{i-1}$ are in the same stage $u_i$ if, and
only if, the values of their parents agree.  This fully expresses the
conditional independence statement embodied in the BN.

Topological characteristics of a CEG derived from a discrete BN
include: \emph{(i)} all the root-to-sink paths have the same length,
\emph{(ii)} the stages consist of situations all of whose distances
(length of the path from the root to the situation) from the root
are the same, and \emph{(iii)} for $2\leq i \leq n$ all stages $u_i$
associated with different configurations of parents of $X_i$ contain
exactly the same number of situations. Examples and d-separation
theorems for CEG's are given in \cite{SmithAndersonCEG}.}
\end{example}

\begin{example} \label{fromCEGtoBN}\normalfont{
Consider the binary BN $X_{2}\leftarrow X_{1}\rightarrow X_{3}$.
Its CEG is in Figure~\ref{fromCEGtoBNfigure}.
Note that the statements that can be read from the topology of this CEG are that the two situations 
$ (\left\{ X_{1}=0,X_{2}=0\right\} ,\left\{ X_{1}=0,X_{2}=1\right\} ) $
and the two situations 
$ (\left\{ X_{1}=1,X_{2}=0\right\} $, $\left\{ X_{1}=1,X_{2}=1\right\} ) $
are in the same stages (respectively $[0,0]\cup \lbrack 0,1]$ and $[1,0]\cup \lbrack 1,1]$).
From the definition of a stage, this means the two conditional statements 
\begin{eqnarray} 
\operatorname{P}(X_{3}=1|\left\{ x_{1}=0,x_{2}=0\right\}) &=& \operatorname{P}(X_{3}=1|\left\{ x_{1}=0,x_{2}=1\right\}) \label{conditone} \\ 
\operatorname{P}(X_{3}=0|\left\{ x_{1}=0,x_{2}=0\right\}) &=& \operatorname{P}(X_{3}=0|\left\{ x_{1}=0,x_{2}=1\right\}) \notag \end{eqnarray} 
and 
\begin{eqnarray} 
\operatorname{P}(X_{3}=1|\left\{ x_{1}=1,x_{2}=0\right\}) &=& \operatorname{P}(X_{3}=1|\left\{ x_{1}=1,x_{2}=1\right\}) \label{condit2} \\ 
\operatorname{P}(X_{3}=0|\left\{ x_{1}=1,x_{2}=0\right\}) &=& \operatorname{P}(X_{3}=0|\left\{ x_{1}=1,x_{2}=1\right\}) \notag \end{eqnarray}
This is synonymous with the statement $X_{3}\indep X_{2}|X_{1}$.
The CEG in Figure~\ref{fromCEGtoBNmodification} contains the statement (\ref{conditone}) but not necessarily (\ref{condit2}), which cannot be expressed using a BN. 
} \end{example}

\section{Manipulation and Causality}\label{SectionSimulators}

\subsection{Manipulations}
A CEG provides a flexible framework for expressing what
might happen were a model to be manipulated in certain ways or made subject to some control.
Of course as Shafer \cite{Shafer} similarly argues for probability trees,
the validity of such a framework is heavily dependent on
context.
Some discussions of notions of the manipulation of a system and
intervention and various applications can be found in
\cite{Hausman98,Pearl2000,Shafer,Spirtesetal93}. Here
we follow Pearl \cite{Pearl2000}: a model for the manipulation
is developed and the issue of suitability of such manipulation for the application under study is left to practical considerations.
Recall briefly the standard definition of ``do''-operator, which is fundamental in \cite{Pearl2000}.
The joint density function of a set of random variables $X_1,\ldots,X_n$ with sample spaces $\mathbb X_1,\ldots, \mathbb X_n$, factorises according to a directed acyclic graph (DAG)
\begin{equation} \label{EqFactDAG}
p(x_1,\ldots,x_n)=\prod_{i=1}^n p(x_i|pa(x_i))
\end{equation}
where $pa(x_i)$ are the parents of $X_i$ in DAG language.
A random variable is forced to assume a certain value with probability one,
say $X_j=\widehat x_j$ for some $j\in\{1,\ldots,n\}$ and $\widehat x_j\in \mathbb X_j$.
A new joint density, $p(\cdot||\widehat x_j)$, is defined on $\{X_1,\ldots,X_n\}\setminus \{X_j\}$ by the formula
\begin{equation} \label{formulaBNmanip}
p(x_1,\ldots,x_{j-1},x_{j+1},\ldots,x_n||\widehat x_j)
=\prod_{i=1,i\neq j}^n p(x_i|\widehat{pa(x_i)})
\end{equation}
where $\widehat{pa(x_i)}$ is the subset of parents of $x_i$ for which $x_j=\widehat x_j$. This formula expresses the effect of the manipulation 
$X_j=\widehat x_j$. 

A manipulation of a probability tree [of a CEG] can be defined in an analogous manner by modifying the distributions of some of the random variables sitting on situations [positions].

\begin{definition} \label{manipulation}
Let $(\mathcal{T},\Pi (\mathcal{T)})$  be a probability tree model and
$D\subset S$ a subset of situations of the tree.
A {\normalfont manipulation} of the tree is a pair
$(D, \widehat{\Pi}_D)$ where
$\widehat{\Pi}_D=\{ \widehat{\pi}(v^\prime|v): v\in D \text{ and } 
v^\prime\in \mathbb{X}(v) \}$
and $\{\widehat{\pi}(v^\prime|v): v^\prime\in \mathbb{X}(v)\}$ is a new
distribution for $X(v)$.

The {\normalfont  effect} of this manipulation is the transformation $
(\mathcal{T},\Pi) \overset{\widehat{\operatorname{P}}}{\rightarrow} (\mathcal{T},\widehat{\Pi
}_D)$ where
\begin{equation} \label{def:manip}
\widehat{\operatorname{P}}({X}(v)=v^\prime)=\left\{ \begin{array}{lr}
\pi(v^\prime|v) & \text{ if } v\notin D \\
\widehat{ \pi}(v^\prime|v) & \text{ if } v\in D
\end{array} \right. \end{equation}
for $v^\prime\in \mathbb{X}(v)$.
The {\normalfont manipulated tree} is the probability tree model so obtained.
The {\normalfont manipulated CEG} is the CEG of the manipulated tree.
\end{definition}

Definition~\ref{manipulation} allows large classes of intervention, some of which are illustrated in Example~\ref{simpletreeInter}.
\begin{example}[Continuation of Example~\ref{simpletree}] \label{simpletreeInter} \normalfont{
(1.) Fix some values of the primitive probabilities or of a function of them, e.g. $\pi_1=1$, $\pi_3+\pi_4=0.5$, $p_4=\pi_1\pi_4=0.5$.
(2.) Impose (polynomial) constraints on primitive probabilities, e.g. $\pi_3=2\pi_4$, $\pi_1=\pi_3=\pi_6$.
(3.) Assume that the distribution of the random variable sitting on some situation is from a parametric family, e.g.
    $X(v_1)$ follows a Binomial distribution with $\operatorname{P}(X(v_1)=v_3)=s^2$, $\operatorname{P}(X(v_1)=v_4)=2s(1-s)$ and $\operatorname{P}(X(v_1)=v_5)=(1-s)^2$ for $s\in [0,1]$.
(4.) In the idle tree  $X(v_1)$ follows a Binomial distribution with $\pi_3=s^2$ and $\pi_5=(1-s)^2$ and
in the manipulated tree $X(v_1)$ has a uniform distribution with $\pi_3=\pi_4=\pi_5=1/3$.
} \end{example}
Here the distinction between intervention and constraint is not a mathematical one. A probability tree model
is assigned together with extra information like its stages or some logical constraints (sum-to-one) or experimental regimes
(women are randomly assigned to treatment B or C in Example~\ref{ABCexample}). These pieces of information impose some constraints on the primitive probabilities.
Then, the system is modified by changing one or more of the distribution of the random variables on situation.
There might be no values of the primitive probabilities that satisfy both the manipulated tree and the idle tree, as in the case of Example~\ref{simpletreeInter} (4). Next we define classes of CEG models and of manipulations for which it is plausible to investigate whether the set of values of the primitive probabilities that satisfy both the idle and the manipulated CEG is not empty.

If two unmanipulated situations were in the same stage of the original tree and are manipulated in the same way, then they remain in the same
stage in the manipulated tree.

\begin{definition}
A manipulation is
called \emph{positioned} if the partition
of positions after the manipulation is equal to or a coarsening of
the partition before manipulation. It is called
\emph{staged} if the partition of stages
after the manipulation is equal to or a coarsening of
the partition before manipulation.
\end{definition}

\begin{example}[Continuation of Example~\ref{simpledagCont}]
\label{Examplesimpledagmanipulated}{\normalfont
Let $v_1$ and $v_2$ be in the same stage with $v_3$ mapping into $v_5$.
The manipulation $D=\{v_1,v_2\}$ and $\widehat{\operatorname{P}}(X(v_1)=v_3)=1$,
$\widehat{\operatorname{P}}(X(v_2)=v_5)=1$ is a staged and a positioned manipulation.
The CEG of the manipulated tree is in Figure~\ref{simpledagmanipulated}
where edges with zero probabilities are not drawn.
} \end{example}

\begin{example}\label{ManipulatedFAX}{\normalfont
In the probability tree in Figure~\ref{fax.eps} the staged manipulation defined by
$\widehat{\pi}(v_5|v_1)= 1$,
$\widehat{\pi}(v_9|v_3)= 1$, 
$\widehat{\pi}(v_{17}|v_{13})= 1$, 
$\widehat{\pi}(v_{19}|v_{17})= 1$, 
leads to a modification of the CEG in Figure~\ref{CEGfax.eps} in which the
edges into $w_\infty$ from the positions $[v_1,v_3]$, $[v_{13}]$ and
$[v_{17}]$ could be not drawn because the associated manipulated probabilities
become zero. Indeed this manipulation could be performed directly on the CEG by removing
the three aforementioned edges. This idea is developed in Section~\ref{manipulatingCEGexample}.
}\end{example}

\begin{example}[Continuation of Example~\ref{ManipulatedFAX}]
\label{ManipulatedFAXbis}{\normalfont
The staged and positioned manipulation with $D=\{v_2,v_7\}$ and
$\widehat{\pi}(v_6|v_2)= 1$, $\widehat{\pi}(v_{12}|v_{7})= 1$ has the effect of cutting off the branch starting at $v_2$ and going through $v_7$. Figure \ref{ManipulatedFAX.Picture} gives the CEG of the manipulated tree.
} \end{example}

A positioned manipulation manipulates all sample units identically
when their future development distributions are identical, using the
same (possibly randomising) allocation rule. A staged manipulation
will treat sample units identically if their next development in the
idle system is the same. Our experience has been that it is often 
sufficient to restrict study to positioned manipulations. We note for 
example that all manipulations on a BN considered by Pearl are positioned and also staged. 
Example \ref{univexample} gives a simple case when a
staged manipulation is not appropriate but a positioned manipulation is.

\begin{example}\label{univexample} \normalfont{
An English university has residence blocks of apartments with two rooms
each. It allocates prospective second year students (either English
(E) or Chinese (C)) to one of the two rooms of each apartment.  The second
room has to be allocated to a prospective first year student. In the
past this has been done at random: that is exactly one of the $N$ second year students to go into an apartment and exactly one
of the $N$ first year students is allocated the integer $1\leq i \leq N$ using a randomization devise and students
share with the student allocated the same integer. However it has been noticed in a
survey that the probability of satisfaction of home students placed
with home students is higher and of Chinese students placed with
Chinese students is higher than when they are mixed. In order to cause
students' satisfaction to increase, the university decides to place
first year students with a second year student with the same
ethnicity.

The BN and CEG of this problem are given in Figure
\ref{University.eps} where $X$ represents the ethnicity of the second
year student, $Y$ that of the first year student and $Z$ is a binary
index of the satisfaction of two students in the same apartment, taking
values $U$ and $S$. Thus for example $X(v_0)=X$, $X(v_1)=[Y|X=E]$,
$X(v_3)=[Z|X=E,Y=E]$ and $\pi(v_5|v_2)$ gives the probability of
allocating a Chinese first year student to an apartment with a Chinese
second year student.  The vertices $v_3$ and $v_5$ are in the same
stage to indicate a non-mixed apartment, analogous interpretation has the
stage $\{v_4,v_6\}$. The undirected edge between $v_1$ and $v_2$
represents the random allocation of the first year student to an apartment.

The relationship between satisfaction and shared race is not depicted
in the BN whilst it is in the CEG through the colouring of its edges.
More significantly it is impossible to determine, either from the
semantics of the BN or the factorisation of the probability mass
function of the path events, whether the allocation of the prospective
second year student occurs before the allocation of the prospective
first year student.  The CEG states that second year allocation occurs
before first year allocation explicitly, so that ``causal''
manipulation of the type suggest by the survey above is a possibility.
The semantic of a BN is not refined enough to represent the sort of manipulation considered in this example.
Note that central to the BN analysis of causal relationships is the absence of edges between vertices, here $X$ and $Y$ (see e.g. \cite{Dawid2002}). The only way to embody the types of manipulation we consider here is to join $X$ and $Y$ by an edge and so loose this intrinsic information.

A manipulation that forces individuals of the same ethnicity to share
an apartment implies a CEG without the directed edge between $v_1$ and $v_2$
and without the crossing arrows in the CEG in Figure
\ref{University.eps}.
}\end{example}

\subsection{Manipulating CEG's} \label{manipulatingCEGexample}
The standard manipulations of a BN are those that force some
components of the network to take pre-assigned values, as in Equation~(\ref{formulaBNmanip}).
The analogue for the CEG is to consider manipulations which force all the
paths to pass through an identified set of positions $W$. For
example the assignment of a particular type of unit, here described by
their current position, to a particular treatment regime, here
described by a set of subsequent positions $W$ (see also Section~\ref{sectioneffectrandomvariable}).

For a CEG $\mathcal C$ and a set of position $W$ in $\mathcal C$, let
$\operatorname{pa}(W)$ denote the set of all parents of the elements
in $W$, that is $\operatorname{pa}(W)= \{ w^\ast \in V(\mathcal C) :
\text{there exists } w\in W \text{ such that } (w^\ast,w) \in
E_d(\mathcal C) \}$.
In the analogy above the set $\operatorname{pa}(W)$
corresponds to the positions any unit must reach to be submitted to a
treatment forcing them into the positions in $W$.

\begin{definition} A subset $W$ of positions of a CEG $\mathcal C$ is
  called a \emph{manipulation set} if 1. all root-to-sink paths in $\mathcal C$ pass through exactly one position in $\operatorname{pa}(W)$, and
  2. each position in $\operatorname{pa}(W)$ has exactly one child in $W$.
\end{definition}

\begin{example}{\normalfont
In the CEG in Figure \ref{simpledagmanipulated} the position
$[v_1,v_2]$ is a manipulation set. Excluding the trivial case of a
manipulation set consisting of the root node only, there is no
manipulation set in the CEG in Figure \ref{CEGfax.eps}.
In the CEG in Figure~\ref{ABCceg} $W=\{[v_1],[v_2]\}$ is not a manipulation set.
The manipulation described in Example \ref{univexample} is to a manipulation set.
} \end{example}

\begin{definition} \label{puremanipulation} A manipulation $(D,
  \widehat{\Pi}_D)$ of a  CEG is
  called a \emph{pure manipulation to the positions} $W$ if
\begin{enumerate}
\item it is a positioned manipulation,
\item for each $v\in D$ there exists $w\in W$ such that
$\operatorname{\widehat{\operatorname{P}}}({X}(v|D)=w)=\widehat{\pi}(w|v)=1$, and
\item no $v\not\in D$ is manipulated.
\end{enumerate}
\end{definition}

A CEG to be \emph{causal} for
an application needs \emph{(i)} to be 
valid for that application and \emph{(ii)} that for the pure
manipulations to any manipulation set the
corresponding manipulated CEG is also valid. If a CEG admits a
description as a BN and the CEG is causal then the BN is also causal
in the sense of \cite[Definition 1.3.1]{Pearl2000}. In this sense a
causal CEG is a natural generalisation of a causal BN, applicable to
asymmetric models. However it can express a larger variety of
manipulations than a causal BN: for example those based on certain
functions of preceding variables as in Example \ref{univexample}. 
Next, three manipulation on the CEG for Example~\ref{BrickBN} are discussed. 

\begin{example}[Continuation of Example~\ref{BrickBN}]\label{brickMan}
\normalfont{\textbf{Manipulation 1.}
Consider the manipulation forced to $w_{1}$---that can be read as ensuring the suspect is taken to court. 
This would assign probability $1$ to the edge labelled $x_{1}$, all vertices other than $w_{\infty }$ on paths after $w_{2}$ and their associated edges are deleted, and the primitive probabilities of the manipulated CEG are like in the unmanipulated CEG except the edge labelled $x_{1}$.
The manipulated CEG is in Figure~\ref{BrickCEGfigureMan1}. 
Notice here that the hypothesis that this new CEG is valid for the manipulated is a substantive one and in particular will depend upon the how we plan to implement the manipulation, for example if whether the suspect went to court simply depended on whether a judge could be found or on government policy. But if we choose not to proceed with prosecution because, on the basis of their evidence, a third party did not think the case was strong enough to convince a jury, then this causal deduction would almost certainly not be valid. 
\textbf{Manipulation 2.} Consider the manipulation forced to $w_{15}$, signifying e.g. that the witness does not identify the suspect. The associated causal CEG is given in Figure~\ref{BrickCEGfigureMan1}. Note that the inevitable consequence of this manipulation is that suspect is released as there is just the release edge $R$ into $w_{\infty}$.  
} \end{example}

\section{Identifying effects of a manipulation}\label{sectionBDtheorem}

\subsection{Identification of causal effects} \label{sectionlemmas}

Recent papers on causal BN literature \cite{Dawid2002,Pearl2000,Pearl2003} study when the
topology of a BN helps to deduce that the effect of a manipulation
on a pre-specified node of the BN can be identified from observing a
subset of the BN variables that are observed or ``manifest'' in an
unmanipulated system. Experiments on the original ``idle'' system can
then be designed so that the effects of, for example, a proposed new
treatment regime, on a manipulated system can be established. Here
we demonstrate that the topology of the CEG can also be
used to find \emph{functions} of the data, for example subsets of
possible measurements, that when observed in the idle system allows
us to estimate effects of a given manipulation. We
first need some definitions.

\subsubsection{Random variables}
Consider $(\mathcal T,\Pi(\mathcal T))$ with sample space $\mathbb X$. 
Any function $Y:\mathbb X\longrightarrow \mathbb R$ is a random variable on $(\mathbb X, 2^\mathbb X, \operatorname{P})$ and defines a partition of the atomic event set, namely
\begin{equation} \label{partitionRV}
\Lambda_y=\left\{ \lambda\in \mathbb X \text{ such that } Y(\lambda)=y
\right\}
\end{equation}
where $y$ ranges over the image of $Y$, written as $\operatorname{Image}(Y)$, which is a finite set. Similarly, given a partition of $\mathbb X$, a random variable $Y$ with finite range space can be constructed so that (\ref{partitionRV}) holds.

Now consider a CEG $\mathcal C$ constructed from the tree. Let $\mathbb X_{\mathcal C}$ be the set of root-to-sink paths in the CEG formed by directed edges. As already mentioned, $\mathbb X_{\mathcal C}$ can be identified with $\mathbb X$ and the $\sigma$-algebra $2^\mathbb X$ on the tree is mapped by the CEG construction into the power set of $\mathbb X_{\mathcal C}$, call it $2^{\mathbb X_{\mathcal C}}$.
Furthermore, the random variable $Y$ corresponds to a random variable on
$(\mathbb X_{\mathcal C}, 2^{\mathbb X_{\mathcal C}})$
that induces on $\mathbb X_{\mathcal C}$  the same partition as the one obtained by mapping the $\Lambda_y$ sets, $y\in \operatorname{Image}(Y)$, on the CEG. Thus, with a slight abuse of notation, 
for $y\in \operatorname{Image}(Y)$ we can write
$\Lambda_y=
\left\{ \lambda\in \mathbb X \text{ such that } Y(\lambda)=y
\right\} =
\left\{ \lambda\in \mathbb X_{\mathcal C} \text{ such that } Y(\lambda)=y
\right\}
$.

\begin{definition} A random variable $Y$ on $\mathbb X$ is called {\normalfont
    observed} (or {\normalfont manifest}) if, and only if, indicators of the
  events $\Lambda_y$ are observed or observable for all $y\in \operatorname{Image}(Y)$.
\end{definition}

Although in practice vectors of manifest random variables are likely to occur, here, without loss of generality, we can work with random variables, because a partition induced by a random vector can be induced by a uni-dimensional random variable.

\subsection{Manipulation forced to a position}
\label{sectioneffectrandomvariable}
\begin{definition}
Call a manipulation of a CEG $\mathcal C$ {\normalfont forced to the position $w$} if after manipulation
\begin{enumerate}
\item  the probability of the event
    $\{w\}=\{\lambda\in\mathbb X: w\in \lambda\}$ is one, and
\item all primitive probabilities in the manipulated CEG associated with positions at or after $w$ in the original CEG are those in the idle system.
\end{enumerate}
\end{definition}

\begin{example}{\normalfont
If $\{w\}$ is a manipulation set of $\mathcal{C}$ then the pure manipulation to $w$ is a manipulation forced to $w$.
 }\end{example}

\begin{example}{\normalfont In Example \ref{univexample} a
    manipulation forced to $w=\{v_3,v_5\}$ is  obtained by setting
    $\widehat\pi(v_4|v_1)=0=\widehat\pi(v_6|v_2)$, that is by allocating
    students with the same ethnicity to the same apartment. This
    manipulation directly on the CEG is given by
    $\widehat\pi(\{v_4,v_6\}|\{v_1\})=\widehat\pi(\{v_3,v_5\}|\{v_2\})=0$.
 }\end{example}

\begin{example}{\normalfont In Example~\ref{contABCexample} the manipulation $\pi_7=1$, forcing a woman who wants to have a child to take medicine $C$, is trivially a positioned manipulation, but not a staged manipulation. It is also a forced manipulation to $w=\{v_7\}$.
}\end{example}

Consider a position $w$ in the CEG $\mathcal C$.
Every directed root-to-sink path through $ w$ can be split into two parts:
one from root to $w$ and one from $w$ to the sink node, $w_\infty$.
Thus for $\lambda\in \mathbb X_\mathcal C$ if $ w\in \lambda $ we can write
$\lambda=\lambda(w_0,w)\times \lambda(w,w_\infty) $,
where $\times$ indicates concatenation of paths.
Note that $\{\lambda(w,w_\infty): \lambda \in \mathbb X_\mathcal C\}$ is the set of root-to-sink paths in the sub-CEG of $\mathcal C$ starting at $w$, namely the CEG whose root is $w$, whose vertex set $V$ is formed by positions in $\mathcal C$ lying on paths from $w$ to $w_\infty$ and whose edges are those in $\mathcal C$ connecting elements in $V$, likewise its edge labels. Call it $\mathcal C(w)$.

Consider a random variable $\widehat Y(w)$ on $(\mathcal C(w), 2^{\mathcal C(w)})$
and let zero be one of the values not taken by  $\widehat Y(w)$.
Let $\{\Lambda^+_y(w) : y\in \operatorname{Image}(\widehat Y(w)) \}$ be the  partition induced by $\widehat Y(w)$ on $\{\lambda(w,w_\infty): \lambda\in \mathbb X_\mathcal C\}$. It
 can be extended to a partition of $\mathbb X_\mathcal C$ with sets
$ \Lambda_0(w)=\{ \lambda\in \mathbb X_\mathcal C : w \not \in \lambda\}$ and 
$ \Lambda_y(w)= \{\lambda=\lambda(w_0,w)\times \lambda(w,w_\infty) : \lambda(w,w_\infty)\in \Lambda^+_y(w) \}
$
for $y\in \operatorname{Image}(\widehat Y(w))$.
Recall that there exists one, usually many, random variables $Y(w)$ on $(\mathbb X_\mathcal C, 2^{\mathbb X_\mathcal C})$ which induce this partition.
In Lemma~\ref{cond=dolemma} we compare the distributions of any such $Y(w)$ and of $\widehat Y(w)$ before and after a manipulation forced to $w$.

\begin{lemma} \label{cond=dolemma}
Consider a manipulation forced to $w$ in the CEG $(\mathcal C,\Pi)$.
Let $\operatorname{P}$ be the probability measure on the unmanipulated CEG and $\widehat{ \operatorname{P}}$ the probability measure on the manipulated CEG,
$(\mathcal C,\widehat \Pi)$.
Let $\{w\}$ be the event of passing through $w$ in the idle system. Let $Y(w)$ and $\widehat Y(w)$ be defined as above.
Then for $y\in \operatorname{Image}(\widehat{Y}(w))$
\begin{enumerate}
\item $\operatorname{P}(\widehat Y(w)=y)= \widehat{\operatorname{P}}(\widehat Y(w)=y)$,
\item $\operatorname{P}(Y(w)=0) = 1- \operatorname{P}(\{w\})$ and
$\widehat{\operatorname{P}}(Y(w)=0) = 0$,
\item $\operatorname{P}(Y(w)=y)= \operatorname{P}(\{w\}) \operatorname{P}(\widehat Y(w)=y)$, and
\item $\widehat{\operatorname{P}}(Y(w)=y)=  \operatorname{P}(\widehat Y(w)=y)$.
\end{enumerate}
\end{lemma}
\begin{proof}
\begin{enumerate}
\item This follows from the fact that the primitive probabilities on $\mathcal C(w)$ are not changed by the manipulation.
\item The probability of not passing through $w$ in the manipulated system is zero because the manipulation is forced to $w$ while in the manipulated system it is clearly one minus the probability of passing through $w$.
\item This is by construction of $Y(w)$ from $\widehat Y(w)$. A path
 $\lambda$ through $w$ is decomposed as $\lambda=\lambda(w_0,w)\times \lambda(w,w_\infty)$. 
By Equation~(\ref{probpaths}) its probability in the idle system is
 $\operatorname{P}(\lambda)=\operatorname{P}(\lambda(w_0,w)) \operatorname{P}(\lambda(w,w_\infty))$. Now, $\operatorname{P}(\lambda(w_0,w))$ is the probability of reaching $w$ and
$\operatorname{P}(\lambda(w,w_\infty))= \operatorname{P}(\widehat Y(w)=y)$ by construction of $Y(w)$ from $\widehat Y(w)$ and the fact that the manipulation forced to $w$ does not change the primitives probabilities for edges after $w$.
\item In the manipulated system $ \widehat{\operatorname{P}}(\{w\})=1$ and use Item~1.
\end{enumerate}
\end{proof}
Item 3 in Lemma~\ref{cond=dolemma} can be re-written as a conditional probability
$$\operatorname{P}(Y(w)=y |\{ w\})=\displaystyle\frac{\operatorname{P}(Y(w)=y)}{\operatorname{P}(\{w\})}=
 \operatorname{P}(\widehat Y(w)=y)  .  $$
Lemma~\ref{cond=dolemma} can be applied if there is a position $w$ such that, after enacting a manipulation, all paths pass through $w$ and 
if the fact that $\{w\}$ occurs can be learnt from a set of measurements in the unmanipulated CEG. This can be checked from the CEG topology.

\subsection{Manipulation forced to a set of positions: the sub-CEG $\mathcal C(W)$} \label{subCEGsection}
It is not always possible, even in models that can be described by a
causal BN, to observe indicators on the events $\{\Lambda_y(w): y\in
\operatorname{Image}(Y) \}$ for a suitable choice of $w$ and $\operatorname{Image}(Y)$.
Nevertheless being able to observe indicators of the set of coarser
events
$\Lambda_y(W)=\bigcup\limits_{w\in W}\Lambda_y (w)
$
for a set of positions, $W$, can also be sufficient for identifiability. To show this is less straightforward although the general set-up is a generalisation of Section~\ref{sectioneffectrandomvariable}.

\begin{definition} \label{regular} A set of positions $W$ of a CEG
  $\mathcal C$ is called {\normalfont $\mathcal{C}$-regular} if no two
  positions in $W$ lie on the same directed path of $\mathcal C$.
\end{definition}
\begin{example} {\normalfont
By definition, a manipulation set of $\mathcal{C}$ is always
$\mathcal{C}$-regular.
}\end{example}

The analogue of the sub-CEG starting at $w$ for a $\mathcal{C}$-regular set of positions, $W$, is a new CEG constructed by joining the sub-CEG's starting at each $w\in W$ to a new root-vertex $w_0^\ast$. The new edge $(w_0^\ast,w)$ is labelled
$
\operatorname{P}(X(w_{0}^{\ast })=w)=\frac{\operatorname{P}(\{w\})}{\operatorname{P}(W)}
$, 
where, as before, $\operatorname{P} (\{ w \} )=\sum_{\lambda\in \mathbb X: w\in
\lambda}\operatorname{P}(\lambda)$ is the probability of passing through $w$ in the
original CEG and $\operatorname{P} (W)=\sum_{w\in W}\sum_{\lambda\in
\mathbb X:w\in \lambda}\operatorname{P}(\lambda)$ is the probability of passing
through a position in the set $W$ in the original CEG. Note that
because $W$ is $C$-regular,
$ \sum_{w\in W}\operatorname{P}(X(w_{0}^{\ast })=w)=1
$
Let $\mathcal C(W)$ be this new CEG.

\begin{example} \normalfont{The manipulation in Example~\ref{ManipulatedFAX} is a manipulation forced to the set of positions
 $W=\{\{v_1,v_3\},\{v_2\}\}$ and $w_0^\ast=\{v_0\}$. A manipulation forced to $W=\{\{v_1,v_3\},\{v_7\}\}$ requires a new $w_0^\ast$ and the edge
  $(w_0^\ast, \{v_7\})$ has the same probability as the edge $(\{v_0\},\{v_2\})$ in the original CEG.
}\end{example}

It can be shown in analogy to
Section~\ref{sectioneffectrandomvariable} that the partition induced
by a random variable $\widehat Y(W)$ on $(
\mathbb{X}^{\mathcal C(W)},2^{\mathbb{X}_{\mathcal C(W)}})$ 
can be extended to a partition on $\mathcal C$ and this, in
turn, can be interpreted as the range space of a random variable on
$(\mathbb X^ \mathcal C,2^{\mathbb X_\mathcal C})$.

\subsection{Manipulation forced to a set of positions: amenable positions}

Let $\mathcal{C}^{\ast }(w)$ denote a graph representing what
happens until we reach a given position $w$. Its vertices and edges
are those along the root-to-$w$ paths in $\mathcal{C}$ and its edge
probabilities are inherited from $ \mathcal{C}$ as well. The graph
$\mathcal{C}^{\ast}(w)$ is not necessarily a CEG, for example in
Figure~\ref{CEGfax.eps} the graph $C^\ast([v_7])$ is not a CEG
because of the undirected edge between $\{v_2\}$ and $\{v_7\}$.
Write $K(\mathcal{C}^{\ast }(w))$ for the set of positions in
$\mathcal{C}$ whose vertices are in $\mathcal{C}^{\ast }(w)$
excluding $w$. For any $\mathcal C$-regular set of positions, $W$,
let $ K(\mathcal{C}^{\ast }(W))=\bigcup\limits_{w \in W
}K(\mathcal{C}^{\ast }(w)) $.

\begin{definition} \label{simpleDef}
Call a set of positions, $W$, \emph{simple} if
\begin{enumerate}
\item $W$ is $\mathcal C$-regular,
\item there exists a partition of $K(\mathcal{C}^{\ast }(W))$
  into $K^{\alpha}(\mathcal{C}^{\ast }(W))$ and
  $K^{\beta}(\mathcal{C}^{\ast }(W))$ called \emph{active} and
  \emph{background} positions respectively such that
\begin{enumerate}
\item 
for each $w\in W$ an \emph{active position} has exactly one emanating
edge along each root-to-$w$ path in $\mathcal C^{\ast }(w)$ if it lies on that path. 
Furthermore for any two positions $w_{1}$,$w_{2}\in W$ 
every pair of root-to-$w_{1}$ path and root-to-$w_{2}$ path containing the
ordered sequences of active positions $w_{1}^{\alpha ,1},w_{1}^{\alpha ,2},\ldots
w_{1}^{\alpha ,n}$ and  $w_{2}^{\alpha ,1},w_{2}^{\alpha
,2},\ldots w_{2}^{\alpha ,n}$ respectively, the pairs of
position $\left\{ \left( w_{1}^{\alpha ,k},w_{2}^{\alpha ,k}\right) :1\leq
k\leq n\right\} $ are in the same stage. If two active positions lie in the
same stage in $\mathcal C^{\ast }(w)$ then their unique emanating edge is
labelled by the probability of the same edge in $\mathcal C$.  

 
\item Each \emph{background position} in $\mathcal C^{\ast }(w)$ inherits a complete
set of emanating edges from $\mathcal C$. Furthermore for any two positions $w_{1}$,
$w_{2}\in W$ every pair of root-to-$w_{1}$ path and root-to-$w_{2}$ path containing the 
ordered sequence of background positions $w_{1}^{\beta,1},w_{1}^{\beta ,2},\ldots w_{1}^{\beta ,n}$ 
and $w_{2}^{1},w_{2}^{2},\ldots w_{2}^{n}$ respectively,
the pairs of position $\left\{ \left( w_{1}^{\beta ,k},w_{2}^{\beta
,k}\right) :1\leq k\leq n\right\} $ are in the same stage. 

\end{enumerate}
\end{enumerate}
\end{definition}


Note that active positions act as labels for the elements of $W$ because each active position is in at most one path through $w\in W$.
Root-to-leaf paths through background positions are governed by the same probability law regardless of the index of $w\in W$. 

\begin{example} \normalfont{
A CEG of the binary CBN given by
\begin{equation*} \begin{array}{rcrclcl}
B &\rightarrow &X &\longrightarrow & Y & \leftarrow &A 
\end{array} \end{equation*}
where $X$ is manipulated to a value $1$ is given by
\begin{equation*}
\begin{array}{ccccccccc}
&  &  & B=1 &  & X=1 &  &  &  \\ 
& A=1 & w_{1}^{\beta ,1} & \rightrightarrows  & w_{1}^{\alpha ,2} & 
\longrightarrow  & w_{1} &  &  \\ 
& \nearrow  & | & B=0 & | &  &  & \searrow ^{\searrow } &  \\ 
w_{0}=w^{\alpha ,1} &  & | &  & | &  &  &  & w_{\infty } \\ 
& \searrow  & | & B=1 & | &  &  & \nearrow \nearrow  &  \\ 
& A=0 & w_{2}^{\beta ,2} & \rightrightarrows  & w_{2}^{\alpha ,2} & 
\rightarrow  & w_{2} &  &  \\ 
&  &  & B=0 &  & X=1 &  &  & 
\end{array}%
\end{equation*}
where the  manipulation set is $W=\left\{ w_{1},w_{2}\right\} $. 
The active and background positions can be identified from $\mathcal C(W)$ thinking of 
$\mathcal C(W)$ as a subgraph of $\mathcal C$ with edge probabilities inherited from $\mathcal C$.
In this example active positions are $\left\{ w_{0},w_{1}^{\alpha ,2},w_{2}^{\alpha
,2}\right\} $ where $\left\{ w_{0},w_{1}^{\alpha ,2}\right\} =\{A=1\}$ and $
\left\{ w_{0},,w_{2}^{\alpha ,2}\right\} =\{A=0\}$ label the two positions
in $W$ and $w_{1}^{\alpha ,2},w_{2}^{\alpha ,2}$ which lie at the same
point in the ordered sequence to $w_{1}$ and $w_{2}$ lie on the same point
in their respective root to $w_{1}$ and root to $w_{2}$ sequence and so lie
in the same position. The background positions are $w_{1}^{\beta ,1}$ and $
w_{2}^{\beta ,1}$, retain all their emanating edges from the original CEG
and are also in the same position. The terminology of active and background position 
here means that active positions might subsequently effect $Y$ if manipulated after $X$ while 
subsequently manipulating background positions cannot effect $Y$.


Further examples are in Section~\ref{BDTsection}. 
}\end{example}

\begin{definition} \label{amenableDef}
A manipulation is called {\normalfont amenable forcing to a set $W$}
if

\begin{enumerate}
\item the set $W$ is simple in $(\mathcal{C},\Pi (\mathcal{C}))$,

\item the set $W$ is simple in $(\mathcal{C},\widehat{\Pi }(\mathcal{C}))$
and $\widehat{\operatorname{P}}(W)=1$, and

\item $\Pi (\mathcal{C})$ and $\widehat{\Pi }(\mathcal{C})$ differ only on
edges whose parents lie in $K^{\alpha }({\mathcal{C}}^{\ast }(W)) $.
\end{enumerate}
\end{definition}
Item 2 in Definition~\ref{amenableDef} assumes that the manipulation
is forced to $W$ as the probability of passing through a vertex not
in $W$ in the manipulated system is zero. Furthermore, an amenable
manipulation may change probabilities on edges associated to active
positions, but will always leave probabilities associated with
background positions unchanged.

\begin{example} \normalfont{
When $W=\{w\}$ is a singleton, the set of active positions will be
empty and so all the conditions above are vacuous and $W$ is
simple. It follows that a pure manipulation forced to $w$ is
amenable.
}\end{example}

\begin{lemma} \label{Lemma4.2}Consider an amenable manipulation forcing to a simple set $W$. Then  for $w\in W$
\[ 
\operatorname{P}(\{w\}) = \pi _{W}^{\alpha }\, \, \operatorname{P}^\beta(\{w\}) \quad \text{ and } \quad 
\widehat{\operatorname{P}}(\{w\}) = \widehat{\pi}_{W}^{\alpha }\, \, \operatorname{P}^\beta(\{w\}) 
\]
where $\operatorname{P}(\{w\})$ is the probability  that a path in $\mathcal C$ will pass through the position $w\in W$ in the idle $(\mathcal{C},\Pi ( \mathcal{C}))$ and $\widehat{\operatorname{P}}(\{w\})$ is the analogue probability in the manipulated system;
$\pi _{W}^{\alpha }$ and
$\operatorname{P}^\beta(\{w\})$
are products of primitive probabilities in $\Pi
(\mathcal{C)}$ associated with random variables whose positions lie in
$K^{\alpha }(\emph{C}^{\ast }(W))$ and $K^{\beta }(\emph{C}^{\ast
}(W))$, respectively.
\end{lemma}
\begin{proof} Item 2 in Definition~\ref{simpleDef} means that events associated with background positions and with active positions are independent. Thus, for each $w\in W$ we have
$
\operatorname{P}(\{w\}) = \operatorname{P}^\alpha(\{w\}) \operatorname{P}^\beta(\{w\})
$
where $\operatorname{P}^\beta(\{w\})$ is defined above and  $\operatorname{P}^\alpha(\{w\})$ is the product of primitive probabilities in $\Pi(\mathcal{C)}$ associated with random variables whose positions lie in $K^{\alpha }(\emph{C}^{\ast }(W))$.
Furthermore, from the definition of $K^{\alpha}(\emph{C}^{\ast }(W))$ for any positions $w,w^\prime\in W$ we have
$
\operatorname{P}^\alpha(\{w\})=\operatorname{P}^\alpha(\{w^\prime\})
=
\pi_{W}^{\alpha }$ (say).

The fact that $W$ is also simple in $(\mathcal{C},\widehat{\Pi }(\mathcal{C}))$ for the amenable manipulation implies that
$
\widehat{\operatorname{P}}(\{w\}) = \widehat{\pi }_{W}^{\alpha }
\widehat{\operatorname{P}}^\beta(\{w\})
$
for all $w\in W$.
Finally from Item~1 in Definition~\ref{amenableDef} we have that
$
\operatorname{P}(\{w\})= \pi _{W}^{\alpha }\, \, \operatorname{P}^\beta(\{w\})
$
and from Item~2
$
\widehat{\operatorname{P}}(\{w\})= \widehat{\pi}_{W}^{\alpha }\, \, \operatorname{P}^\beta(\{w\}) $.
\end{proof}

\begin{lemma} \label{amenablelemma}
Consider an amenable manipulation forcing to a simple set $W$.
The distribution of a random variable $\widehat{Y}(W)$ defined on the sub-CEG $\mathcal C(W)$ is identified from the probabilities in the unmanipulated system of the events $\{Y(W)=y,W\}$ for $y\in \operatorname{Image}(Y)$ where $Y(W)$ is constructed as above and
 its probabilities are given by the equation
\begin{equation*}
\widehat{\operatorname{P}}(\widehat{Y}(W)=y|W)=\frac{\operatorname{P}(Y(W)=y|W)}{\operatorname{P}(W)}
\end{equation*}
provided that $\operatorname{P}(\{w\})>0$ for all $w\in W$.
\end{lemma}
\begin{proof} Clearly for $y\in \operatorname{Image}(Y)$
\[
\operatorname{P}(Y(W)=y|W) = \displaystyle \frac{\operatorname{P}(Y(W)=y, W) }{ \operatorname{P}(W) }
\]
where $\operatorname{P}(W)=\sum_{w\in W}  \operatorname{P}(\{w\}) = \pi^\alpha_W \sum_{w\in W}  \operatorname{P}^\beta(\{w\}) $  by Lemma~\ref{Lemma4.2}. Analogously
\[
\widehat{\operatorname{P}}(\widehat{Y}(W)=y|W) = \displaystyle \frac{\widehat{\operatorname{P}}(\widehat{Y}(W)=y) }{ \operatorname{\widehat{\operatorname{P}}}(W) }
\]
where $ \operatorname{\widehat{\operatorname{P}}}(W)=1$ as the manipulation is forced to $W$ and by Lemma~\ref{Lemma4.2}
$ \operatorname{\widehat{P}}(W)=\sum_{w\in W} \operatorname{\widehat{P}}(\{w\}) = \widehat{\pi}^\alpha_W \sum_{w\in W}  \operatorname{P}^\beta(\{w\})$. Furthermore
\begin{eqnarray*}
\widehat{\operatorname{P}}(\widehat{Y}(W)=y) &=&
\sum_{w\in W} \widehat{\operatorname{P}}(\widehat{Y}(W)=y|W) \widehat{\operatorname{P}}(\{w\}) \\
&=& \sum_{w\in W}  \widehat{\pi}^\alpha_W \operatorname{P}^\beta(\{w\}) \widehat{\operatorname{P}}(\widehat{Y}(W)=y|\{w\}) \text{ by Lemma~\ref{Lemma4.2} (8)} \\
&=& \widehat{\pi}^\alpha_W \sum_{w\in W} \operatorname{P}^\beta(\{w\}) \widehat{\operatorname{P}}(\widehat{Y}(W)=y|\{w\}) \text{ multiply and divide by $\pi^\alpha_W$}\\
&=& \displaystyle\frac{\widehat{\pi}^\alpha_W}{\pi^\alpha_W}
\sum_{w\in W} \operatorname{P}^\beta(\{w\}) \pi^\alpha_W \widehat{\operatorname{P}}(\widehat{Y}(W)=y|\{w\}) \\
&=& \displaystyle\frac{\widehat{\pi}^\alpha_W}{\pi^\alpha_W}
\sum_{w\in W} \operatorname{P}(\{w\}) \widehat{\operatorname{P}}(\widehat{Y}(W)=y|\{w\})  .
\end{eqnarray*}
Set $\{\widehat{Y}(W)=y|\{w\}\}=\{\widehat X(w)=y\}$ and use Lemma~\ref{cond=dolemma} Item 1 to obtain
\[\widehat{\operatorname{P}}(\widehat{Y}(W)=y)
=
\displaystyle\frac{\widehat{\pi}^\alpha_W}{\pi^\alpha_W}
\sum_{w\in W} \operatorname{P}(\{w\}) {\operatorname{P}}(\widehat{X}(W)=y) . \]
As for $w\in W$ it holds $\operatorname{P}(\{w\})= \operatorname{P}(\{w\}|W) \operatorname{P}(W)$ then
\begin{eqnarray*}
\widehat{\operatorname{P}}(\widehat{Y}(W)=y) 
=
\displaystyle\frac{\widehat{\pi}^\alpha_W}{\pi^\alpha_W}
\sum_{w\in W} \operatorname{P}(\{w\}|W) {\operatorname{P}}(\widehat{Y}(W)=y|\{w\})  
=
\displaystyle\frac{\widehat{\pi}^\alpha_W}{\pi^\alpha_W} {\operatorname{P}}(\widehat{Y}(W)=y|W)
\end{eqnarray*}
Thus we have that $\widehat{\operatorname{P}}(\widehat{Y}(W)=y)$ is proportional to $ {\operatorname{P}}(\widehat{Y}(W)=y|W)$ and, being probabilities, they must be equal.
\end{proof}

\section{A back-door theorem for CEG's}\label{BDTsection}

In \cite{Pearl2000} sufficient topological conditions (see Definition 3.3.1 page 79) on a causal BN are given for when the probability of a random variable $Y$ on the BN is governed by the
formula
\[
\widehat{\operatorname{P}}_x(Y)=\sum_{z\in \mathbb Z} \operatorname{P}(Y=y| X=x, \mathbf  Z=\mathbf z) \operatorname{P}(\mathbf Z=\mathbf z)
\]
after $X$ has been manipulated to take a value $x$.
Here $\mathbf Z$ is a random vector of variables appearing as vertices in the BN and $\mathbb Z$ is its sample space. A particular consequence of this formula is that
$\widehat{\operatorname{P}}_x(Y)$ can be calculated as a function of the marginal probability distribution of $(X,Y,\mathbf Z)$ and we do not need to observe any other variable in the system.

Next an analogue formula is derived for CEG's.
Let $(Y,Z)$ be a random vector taking values $(y,z)\in \operatorname{Image}(Y)\times \operatorname{Image}(Z)$ and let $W$ be a set of positions in a CEG $\mathcal C$. We will show that it is sufficient to know the probabilities of the events $\{Y=y, Z=z, W \}$. 
As above let $\widehat {\operatorname{P}}(Y(W)=y)$ denote the probability of $Y$ after a manipulation to a set of positions $W$.

For every $z\in \operatorname{Image}(Z)$ let $\Omega_z$ be the set of root-to-leaf paths corresponding to the event $\{Z=z\}$.
Assume that there exists a regular set of positions $W_z$ such that the set of root-to-leaf paths through $W_z$ is equal to $\Omega_z$.
Let $\mathcal C(W_z)$ be  the sub-CEG defined as in Section~\ref{subCEGsection} with new root vertex $w_0(z)$.

\begin{theorem} \label{BDT}
Consider a random variable $Z$ on a CEG $\mathcal C$ and a manipulation forced to a set of positions $W$ in $\mathcal C$.
\begin{itemize}
\item[(i)] For $z\in \operatorname{Image}(Z)$ let $W(z)$ be the set of manipulated positions on root-to-leaf paths passing through a position in $W_z$.
\item[(ii)] Assume that if in $\mathcal C(W_z)$ $w^\prime\in W(z)$ and $w\in W_z$ are on the same root-to-leaf path then $w^\prime$ comes after $w$ in the path ordering.
\item[(iii)] Assume that for each $z\in \operatorname{Image}(Z)$ the manipulation in $\mathcal C(W_z)$ is amenable forcing to $W(z)$.
\end{itemize}
Let $\widehat Y(W)$ be a random variable on the manipulated CEG. For $y\in \operatorname{Image}(\widehat Y(W))$ and $z \in \operatorname{Image}(Z)$ let
$\{ \widehat Y(W_z)=y \}$ be the set of root-to-leaf paths for the event
$\{ \widehat Y(W)=y\}$ intersected with the set of root-to-leaf paths through
$W_z$. As before, let $Y(W)$ be the extension of $ \widehat Y(W)$ to the unmanipulated CEG. Then 
\begin{equation} \label{2}
\widehat{\operatorname{P}}(Y(W)=y) = \sum_{z\in \operatorname{Image}(Z)} \operatorname{P}(Y(W)=y|Z=z)\operatorname{P}(Z=z) . 
\end{equation}
\end{theorem}
\begin{proof} From \emph{(i)} for each $z\in \operatorname{Image}(Z)$ the conditional events
$\{ Y=y,W| Z=z\}=\{ Y=y,W(z)| Z=z \}$, $y\in  \operatorname{Image}(Y)$
are all measurable with respect to the power set of $\mathcal C(W_z)$.
From \emph{(ii)} $\widehat{\operatorname{P}}(Z=z)=\operatorname{P}(Z=z)$,
and by Lemma~\ref{amenablelemma} and condition \emph{(iii)} we have
$ \widehat{\operatorname{P}}(Y(W_z)=y|Z=z)=\operatorname{P}(Y(W_z=y|Z=z) $.
So
\begin{eqnarray*}
\widehat{\operatorname{P}}(Y(W_z)=y)
&=&\widehat{\operatorname{P}}(Y(W_z)=y|Z=z) \widehat{\operatorname{P}}(Z=z)
 \text{ by the definition of $W(z)$} \\
&=& \operatorname{P}(Y(W_z)=y|Z=z) {\operatorname{P}}(Z=z)
 \text{ by the two identities above} \\
&=& \operatorname{P}(Y(W)=y|Z=z) {\operatorname{P}}(Z=z)
 \text{ by the definition of $W(z)$}.
\end{eqnarray*}
\end{proof}
The topology of each graph $\mathcal C(W_z)$ is inherited from
$\mathcal C$ except for the new root vertex $w_0(z)$ and its
connecting edges. So the topological conditions on $\mathcal C(W_z)$
given in Theorem~\ref{BDT} are inherited as topological conditions
on the idle CEG $\mathcal C$. Theorem \ref{BDT} is particularly
useful when the topologies of $\mathcal{C}(W_z)$, $z\in
\operatorname{Image}(Z)$, are different, evoking different
ways of satisfying the criteria of Theorem \ref{BDT} for different
configurations $z$. Of course, if it is possible to express a
model using a BN, then $\mathcal{C}(W_z)$, $z\in
\operatorname{Image}(Z)$ are all identical.

\begin{example}[Continuation of
Example~\ref{univexample}]
\label{univexampleBDT}
\normalfont{ First year students
making the university first
choice $\{Z=0\}$ will be
allocated a shared apartment on
campus whilst others $\{Z=1\}$
will be lodged either in town
$W$, namely $\{X_3=0 \}$, or in
town $C$, namely $\{X_3=1\}$
and have a friendly landlord
$\{U=0\}$ or an unfriendly
landlord $\{ U=1\}$. When $\{
Z=0\}$ it is believed that the
CEG of
Figure~\ref{University.eps} is
valid. If $\{Z=1\}$ the town is
chosen independently of the race
$X_2$ of the first year student,
the friendliness of the landlord
does not depend on the town or
race of the student. However the
satisfaction $Y$ depend both on
friendliness of the landlord and
the allocated town $C$: $C$
having a higher probability of
higher satisfaction $\{ Y=1\}$
than $W$, conditional on
$\{Z=1\}$. This scenario can be
expressed as a BN on four random
variables in
Figure~\ref{aaa}
\begin{equation} \label{aaa}\begin{array}{ccccc}
X_2\quad & U &\longrightarrow Y & \longleftarrow & X_3
\end{array} \end{equation}
We want to consider a proposed manipulation of
our allocation policy for next year. We plan to match campus
students so that those sharing an apartment are of the same race and to allocate off
campus students only to lodgings in town $C$. Our interest is in
$\widehat{P} (Y(W)=1)$ i.e. the overall predicted probability of
high satisfaction were we to implement this policy.
We plan to estimate this probability with a small data set,
collected from earlier years. The sort of asymmetries exhibited by
this problem makes extremely awkward to represent it through a
single BN. Thus $X_1$ is only defined for a student allocated to
campus whilst $(X_3,U)$ only to students allocated to lodgings.
Furthermore the manipulation proposed is different for different
contingencies.

The whole problem can be represented by the CEG in
Figure~\ref{CEGsimpleBDTaaa} where the set of positions manipulated
are coloured in black. Note that $\{Z=0\}$ and $\{Z=1\}$ define a
cut and for our proposed manipulation condition (i) in
Theorem~\ref{BDT} is satisfied. 

The topology of $\mathcal C_{Z=0}$ is identical to the sub-graph of $\mathcal C$
consisting of all edges and vertices on root-to-leaf paths
containing edge $\{Z=0\}$. Similarly, $\mathcal C_{Z=1}$ can be identified with the sub-graph of
$\mathcal C$ consisting of all edges and vertices on root-to-leaf
paths containing the edge $\{Z=1\}$.

Since $W(0)$ is a singleton, applying Lemma \ref{cond=dolemma} to
$\mathcal C_{Z=1}$ gives us that
\[
\widehat{\operatorname{P}}(Y(W(0))=1|Z=0) = \operatorname{P}(Y=1| X=0, Z=0) .
\]
A manipulation to $W(1)$ on 
$\mathcal C_{Z=1}$ is clearly amenable so that
\[
\widehat{\operatorname{P}}(Y(W(0))=1|Z=1) = \operatorname{P}(Y=1| X=1, Z=0) .
\]
It follows that
\[
\widehat{\operatorname{P}}(Y(W(0))=1)
= \operatorname{P}(Y=1| X=0, Z=0) \operatorname{P}(Z=0) 
+ \operatorname{P}(Y=1| X=1, Z=1) \operatorname{P}(Z=1) .
\]

Thus $\widehat{\operatorname{P}}(Y(W)=1)$ is expressed as a function
of only three probabilities from the idle system: the probability
that a student is on campus, the probability that a campus student sharing with someone of the same race 
makes a favourable return given she shares the race of her room-mate
and the probability that a non campus student residing in town $C$ makes
a favourable return. It follows that we have been able to deduce
from the topology of $\mathcal C$ that the probability of the
ethnicity of match pairs of campus students and the conditional
distribution of returns of unmatched students are irrelevant to
$\widehat P$. Furthermore, the race and probability of friendliness
of the landlord for lodged students is also irrelevant to this
calculation and need not be estimated. Note here that we have
deduced what function of variables is sufficient to discover
$\widehat P$: here $X=|{X_1-X_2}|/2$, a feature that cannot be
deduced directly from any BN on the original measurement variables.
} \end{example}

\begin{corollary} Consider a causal CEG, $Z$ and $Y$ as in Theorem~\ref{BDT} and a pure manipulation to a
  manipulation set $W$. If all the events $
  \{Y=y,W,Z=z\}$, $y\in \operatorname{Image}(Y)$ and  $z\in \operatorname{Image}(Z)$ are
  manifest, then the effect of the manipulation is identified and
  given by Equation~(\ref{2}) whenever $W$ is simple, conditioned
  on $Z$.
\end{corollary}
\begin{proof} It follows directly from Theorem~\ref{BDT}. \end{proof}

\begin{example} \normalfont{
Note that if a CEG of a BN is constructed so that the back-door
variables are introduced as early as possible compatibly with the
ordering of the BN, then the conditions of Theorem~\ref{BDT} are
satisfied for atomic interventions on a causal BN. }\end{example}

\begin{example}[Continuation of Example~\ref{brickMan}] \label{BDforbrick} \normalfont{\textbf{Manipulation 3.} 
Consider a third manipulation forced to the set of positions $W=\{w_{14},w_{16}\}$---i.e. the witness is forced to positively identify the suspect. The CEG $\mathcal{C} (W)$ is given in Figure~\ref{BrickCEGfigureMan3}.
Suppose we learn the values of $Z$ are $1$ if a path passes through $w_{7}$, $2$ if through $w_{8}$ and $3$ if $w_{4}$, meaning respectively that
the suspect threw the brick, the suspect was present but did not throw the brick, the suspect not present. 
The conditions of Theorem~\ref{BDT} are satisfied. For example conditioning on $\left\{ Z=1\right\} $ gives us the graph in Figure~\ref{BrickCEGfigureMan3bis}.
In this sub-graph $\{w_{7})$ is the only active position and $\{w_{0},w_{1},w_{3},w_{10},w_{11}\}$ are the background positions.
Similarly it holds for the sub-graphs associated with $\left\{ Z=2\right\} $ and $\left\{ Z=3\right\}$. 
Therefore we can conclude that probability of conviction $\left\{ X_{6}=1\right\}$ after this manipulation is 
\begin{equation*} \operatorname{\widehat{P}}(X_{6}=1)=\sum_{z=1}^{3}\operatorname{P}(Z=z)\operatorname{P}(X_{6}=1|Z=z\text{,}X_{5}=1) \end{equation*}
This probability can be estimated from data on similar cases, provided that in these cases the joint distribution of $\{Z,X_{6}|X_{5}\}$ is known and recorded. 
}\end{example}

\section{Discussion}
Some problems, which are not
satisfactorily expressed in terms of the exchangeable relationships
in a BN (see \cite{RiccoSmith2003,RiccomagnoandS04} and Example~\ref{univexample}),
can be well represented by a CEG model.
This applies when the sample space is asymmetric as in the CEG in Figure~\ref{CEGsimpleBDTaaa} or
the order of factors may be different for different settings of the factors as in
Example~\ref{ABCexample}.

Sometimes as in Example~\ref{univexample}, causality is naturally
expressed through predictions concerning the manipulation of
unfolding situations rather than through assertions about the
effects of manipulations on dependence relationships between
measurements. Example~\ref{univexampleBDT} shows that to determine
the effect of a cause function of the measurements, namely
$|X_1-X_2|/2$,  may be sufficient instead of resorting only to
subsets of measurements like in BN's. BN technology encourages to
express causal hypotheses in terms of the random variables and
parametrization in which data are conveyed to us within a certain
parametrisation. It is now well appreciated that it is often
necessary to separate causal structure from the dependence
structures introduced into measurements through a particular
sampling mechanism specific to the acquisition of information for a
particular study. CEG modelling allows that.

Note that the high number of vertices in a probability tree is
reduced in a CEG by the sink node that collects the leaves and by
the modelling constraints imposed by the position equivalence
relation. However in some cases it can remain much larger than the
number of factors in the modelled problem. For this reason we do not recommend the
use of a CEG model where a simpler structure, like a BN, can be
usefully employed. Furthermore the BN gives a representation of dependence structure that does not depend on the
sample space of random variables whilst the CEG needs this sample space to be specified and be
finite. Sometimes when addressing causal ideas we do not want the size of the nature or size of
the sample space to intrude in which case we are again forced back on to a BN representation.

Another difficulty with CEG's is that to our knowledge, the characterization of the equivalence classes
both associated with CEG's whose probabilistic structure is identical and also those associated
with graphs which can causally be identified are as yet only partially understood. This is
in contrast to BN's where such equivalence classes can be
determined---respectively by the pattern (or essential graph) or the BN itself. For tasks like
model selection a good understanding of such equivalence classes is important to develop for
CEG's. We plan to report on this topic in a future paper. Finally, whilst much more varied
dependence relationships can be expressed using a CEG rather than a BN, CEG
is also limited in the number of factorization formulae it can express simultaneously, being a graph. When
such structure is present the CEG no longer provides a fully topological framework expressing
all the contingent independence relationships and our criticisms of BN's apply equally to
CEG's. We are then thrown on to causal analyses using more algebraic techniques \cite{RiccomagnoSmith2007}.
Despite these caveats we believe that the CEG's provide a powerful graphical tool for
the study of implicit causal relationships derivable from its qualitative structure. 

We now turn to briefly discuss some generalisations. In the paper we
considered the power set $\sigma$-algebra to limit technicalities. But
generalisations of our results can be considered for less refined
$\sigma$-algebras. Searching over functions of measurements to find the cheapest way of
identifying a quantity of interest will often be of much great
value. This will be particularly useful if those measurements have
not yet been collected, or their parametrisations have been chosen
by convention rather than because they reflect in some natural way
the mechanism by which things happen. Search algorithms need to be
developed. 
Example~\ref{simpletree} shows that techniques from algebraic
geometry can be usefully employed on CEG's as they have been on BN's
\cite{GarciaStillmanSturmfels2005} and for identification of causal
effects on BN's \cite{RiccoSmith2003}. This is only in part being
explored and could be combined with search algorithms and design
issues.


\begin{figure}
\setlength{\unitlength}{0.25cm}
\begin{picture}(50,22)(2,-6)\thicklines

\put(0,5){\circle{2}} \put(0,5){\makebox(0,0){$w_{0}$}}

\put(10,5){\circle{2}} \put(10,5){\makebox(0,0){$w_{1}$}}
\put(10,10){\circle{2}} \put(10,10){\makebox(0,0){$w_{3}$}}

\put(20,15){\circle{2}} \put(20,15){\makebox(0,0){$w_{7}$}}
\put(20,10){\circle{2}} \put(20,10){\makebox(0,0){$w_{8}$}}
\put(20,5){\circle{2}} \put(20,5){\makebox(0,0){$w_{4}$}}

\put(30,15){\circle{2}} \put(30,15){\makebox(0,0){$w_{10}$}}
\put(30,10){\circle{2}} \put(30,10){\makebox(0,0){$w_{11}$}}
\put(30,5){\circle{2}} \put(30,5){\makebox(0,0){$w_{12}$}}
\put(30,-5){\circle{2}} \put(30,-5){\makebox(0,0){$w_{13}$}}

\put(40,15){\circle{2}} \put(40,15){\makebox(0,0){$w_{14}$}}
\put(40,5){\circle{2}} \put(40,5){\makebox(0,0){$w_{15}$}}
\put(40,-5){\circle{2}} \put(40,-5){\makebox(0,0){$w_{16}$}}

\put(50,0){\circle{2}} \put(50,0){\makebox(0,0){$w_\infty$}}

\put(1,5){\vector(1,0){8}}   \put(4,4){$x_1$}

\put(10,6){\vector(0,1){3}}   \put(8,7){$x_2$}

\put(11,10){\vector(2,1){8}}  \put(14,13){$x_3$}
\put(11,10){\vector(1,0){8}} \put(14,9){$\bar x_3$}
\put(11,5){\vector(1,0){8}}   \put(14,5.5){$\bar x_2$}

\put(20,6){\line(0,1){3}}   

\put(21,15){\vector(1,0){8}} \put(25,15.5){$x_4$}
\put(21,15){\vector(2,-1){8}} \put(23,14){$\bar x_4$}
\put(21,10){\vector(1,0){8}} \put(26,11){$\bar x_4$}
\put(21,10){\vector(2,1){8}} \put(25,9.5){$x_4$} 
\put(21,5){\vector(1,0){8}} \put(25,5.5){$x_4$}
\put(21,5){\vector(1,-1){9}} \put(25,0){$\bar x_4$}

\put(30,-4){\line(0,1){8}}  
\put(30,11){\line(0,1){3}}  

\put(31,15){\vector(1,0){8}} \put(35,15.5){$x_5$}
\put(31,15){\vector(1,-1){9}} \put(35,11){$\bar x_5$}
\put(31,10){\vector(2,-1){8}} \put(35,8){$\bar x_5$}
\put(31,5){\vector(1,0){8}} \put(35,4){$\bar x_5$} 
\put(31,-5){\vector(1,0){8}} \put(35,-4.5){$\bar x_5$}
\put(31,5){\vector(1,1){9}} \put(32,6){$x_5$} 
\put(31,-5){\vector(1,2){9.5}} \put(32,-4){$x_5$}
\put(31,-5){\vector(1,1){9}} 
\put(31,10){\vector(2,-3){9}} 

\put(41,5){\vector(2,-1){8}} \put(45,3){R}
\qbezier(41.5,15)(48,9)(50,1) \put(50,1){\vector(1,-1){0.1}}  \put(46,10){R}
\qbezier(41.5,15)(44,6)(50,1) \put(50,1){\vector(1,-1){0.1}}  \put(45,6){C}
\qbezier(41,-5)(45,-2)(49,0) \put(49,0){\vector(1,1){0.1}}  \put(45,-1.5){R}
\qbezier(41,-5)(45,-5)(49,0) \put(49,0){\vector(1,1){0.1}}  \put(45,-4.5){C}

\end{picture}
\caption{CEG for Manipulation 1 in Example~\ref{brickMan}}
\label{BrickCEGfigureMan1}
\end{figure}

\begin{figure}
\setlength{\unitlength}{0.25cm}
\begin{picture}(50,24)(2,-6)\thicklines

\put(0,0){\circle{2}} \put(0,0){\makebox(0,0){$w_{0}$}}

\put(10,5){\circle{2}} \put(10,5){\makebox(0,0){$w_{1}$}}
\put(10,10){\circle{2}} \put(10,10){\makebox(0,0){$w_{3}$}}

\put(20,15){\circle{2}} \put(20,15){\makebox(0,0){$w_{7}$}}
\put(20,10){\circle{2}} \put(20,10){\makebox(0,0){$w_{8}$}}
\put(20,5){\circle{2}} \put(20,5){\makebox(0,0){$w_{4}$}}

\put(30,15){\circle{2}} \put(30,15){\makebox(0,0){$w_{10}$}}
\put(30,10){\circle{2}} \put(30,10){\makebox(0,0){$w_{11}$}}
\put(30,5){\circle{2}} \put(30,5){\makebox(0,0){$w_{12}$}}
\put(30,-5){\circle{2}} \put(30,-5){\makebox(0,0){$w_{13}$}}

\put(40,5){\circle{2}} \put(40,5){\makebox(0,0){$w_{15}$}}

\put(50,0){\circle{2}} \put(50,0){\makebox(0,0){$w_\infty$}}

\put(1,0){\vector(2,1){8}}   \put(4,3){$x_1$}

\put(10,6){\vector(0,1){3}}   \put(8,7){$x_2$}

\put(11,10){\vector(2,1){8}}  \put(14,13){$x_3$}
\put(11,10){\vector(1,0){8}} \put(14,9){$\bar x_3$}
\put(11,5){\vector(1,0){8}}   \put(14,5.5){$\bar x_2$}

\put(20,6){\line(0,1){3}}   

\put(21,15){\vector(1,0){8}} \put(25,15.5){$x_4$}
\put(21,15){\vector(2,-1){8}} \put(23,14){$\bar x_4$}
\put(21,10){\vector(1,0){8}} \put(26,11){$\bar x_4$}
\put(21,10){\vector(2,1){8}} \put(25,9.5){$x_4$} 
\put(21,5){\vector(1,0){8}} \put(25,5.5){$x_4$}
\put(21,5){\vector(1,-1){9}} \put(25,0){$\bar x_4$}

\put(30,-4){\line(0,1){8}}  
\put(30,11){\line(0,1){3}}  

\put(31,15){\vector(1,-1){9}} \put(35,11){$\bar x_5$}
\put(31,10){\vector(2,-1){8}} \put(35,8){$\bar x_5$}
\put(31,5){\vector(1,0){8}} \put(35,4){$\bar x_5$} 
\put(31,-5){\vector(1,1){9}} 

\put(41,5){\vector(2,-1){8}} \put(45,3){R}

\end{picture}
\caption{CEG for Manipulation 2 in Example~\ref{brickMan}}
\label{BrickCEGfigureMan2} 
\end{figure}


\begin{figure}
\setlength{\unitlength}{0.25cm}
\begin{picture}(50,22)(2,-6)\thicklines
\put(0,5){\circle{2}} \put(0,5){\makebox(0,0){$w_{0}$}}

\put(10,5){\circle{2}} \put(10,5){\makebox(0,0){$w_{1}$}}
\put(10,10){\circle{2}} \put(10,10){\makebox(0,0){$w_{3}$}}

\put(20,15){\circle{2}} \put(20,15){\makebox(0,0){$w_{7}$}}
\put(20,10){\circle{2}} \put(20,10){\makebox(0,0){$w_{8}$}}
\put(20,5){\circle{2}} \put(20,5){\makebox(0,0){$w_{4}$}}

\put(30,15){\circle{2}} \put(30,15){\makebox(0,0){$w_{10}$}}
\put(30,10){\circle{2}} \put(30,10){\makebox(0,0){$w_{11}$}}
\put(30,5){\circle{2}} \put(30,5){\makebox(0,0){$w_{12}$}}
\put(30,-5){\circle{2}} \put(30,-5){\makebox(0,0){$w_{13}$}}

\put(40,15){\circle{2}} \put(40,15){\makebox(0,0){$w_{14}$}}
\put(40,-5){\circle{2}} \put(40,-5){\makebox(0,0){$w_{16}$}}

\put(50,0){\circle{2}} \put(50,0){\makebox(0,0){$w_\infty$}}

\put(1,5){\vector(1,0){8}}   \put(4,4){$x_1$}

\put(10,6){\vector(0,1){3}}   \put(8,7){$x_2$}

\put(11,10){\vector(2,1){8}}  \put(14,13){$x_3$}
\put(11,10){\vector(1,0){8}} \put(14,9){$\bar x_3$}
\put(11,5){\vector(1,0){8}}   \put(14,5.5){$\bar x_2$}

\put(20,6){\line(0,1){3}}   

\put(21,15){\vector(1,0){8}} \put(25,15.5){$x_4$}
\put(21,15){\vector(2,-1){8}} \put(23,14){$\bar x_4$}
\put(21,10){\vector(1,0){8}} \put(26,11){$\bar x_4$}
\put(21,10){\vector(2,1){8}} \put(25,9.5){$x_4$} 
\put(21,5){\vector(1,0){8}} \put(25,5.5){$x_4$}
\put(21,5){\vector(1,-1){9}} \put(25,0){$\bar x_4$}

\put(30,-4){\line(0,1){8}}  
\put(30,11){\line(0,1){3}}  

\put(31,15){\vector(1,0){8}} \put(35,15.5){$x_5$}
\put(31,-5){\vector(1,0){8}} \put(35,-4.5){$\bar x_5$}

\qbezier(41.5,15)(48,9)(50,1) \put(50,1){\vector(1,-1){0.1}}  \put(46,10){R}
\qbezier(41.5,15)(44,6)(50,1) \put(50,1){\vector(1,-1){0.1}}  \put(45,6){C}
\qbezier(41,-5)(45,-2)(49,0) \put(49,0){\vector(1,1){0.1}}  \put(45,-1.5){R}
\qbezier(41,-5)(45,-5)(49,0) \put(49,0){\vector(1,1){0.1}}  \put(45,-4.5){C}

\end{picture}
\caption{CEG for Manipulation 3 in Example~\ref{brickMan}}
\label{BrickCEGfigureMan3}
\end{figure}


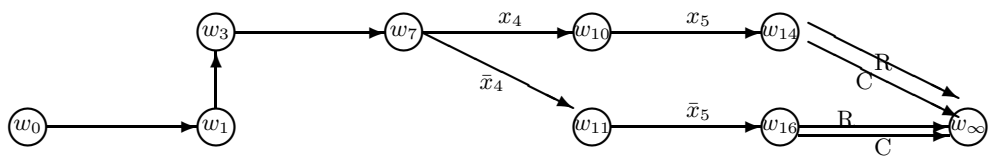
\begin{figure}
\setlength{\unitlength}{0.25cm}
\begin{picture}(50,5)(2,9)\thicklines

\put(0,10){\circle{2}} \put(0,10){\makebox(0,0){$w_{0}$}}

\put(10,10){\circle{2}} \put(10,10){\makebox(0,0){$w_{1}$}}
\put(10,15){\circle{2}} \put(10,15){\makebox(0,0){$w_{3}$}}

\put(20,15){\circle{2}} \put(20,15){\makebox(0,0){$w_{7}$}}

\put(30,15){\circle{2}} \put(30,15){\makebox(0,0){$w_{10}$}}
\put(30,10){\circle{2}} \put(30,10){\makebox(0,0){$w_{11}$}}

\put(40,15){\circle{2}} \put(40,15){\makebox(0,0){$w_{14}$}}
\put(40,10){\circle{2}} \put(40,10){\makebox(0,0){$w_{16}$}}

\put(50,10){\circle{2}} \put(50,10){\makebox(0,0){$w_\infty$}}

\put(1,10){\vector(1,0){8}}  
\put(10,11){\vector(0,1){3}} 

\put(11,15){\vector(1,0){8}} 

\put(21,15){\vector(1,0){8}} \put(25,15.5){$x_4$}
\put(21,15){\vector(2,-1){8}} \put(24,12){$\bar x_4$}

\put(31,15){\vector(1,0){8}} \put(35,15.5){$x_5$}
\put(31,10){\vector(1,0){8}} \put(35,10.5){$\bar x_5$}

\put(41.5,15.5){\vector(2,-1){8}} \put(45,13){R}
\put(41.5,14.5){\vector(2,-1){8}} \put(44,12){C}
\put(41,10){\vector(1,0){8}} \put(43,10){R}
\put(41,9.5){\vector(1,0){8}}  \put(45,8.5){C}

\end{picture}
\caption{Backdoor theorem for Example~\ref{BDforbrick}}
\label{BrickCEGfigureMan3bis}
\end{figure}


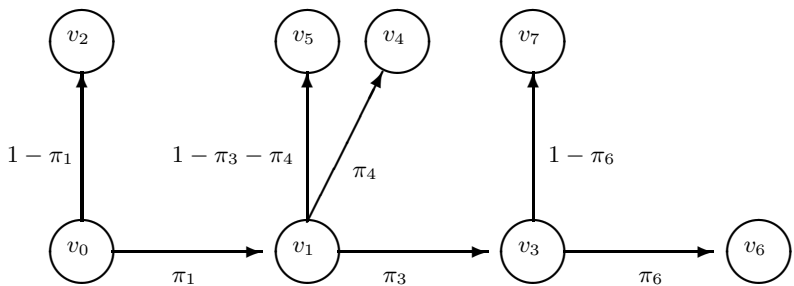
\begin{figure}
\setlength{\unitlength}{0.2cm}
\begin{picture}(40,20)(1,-1)\thicklines
\put(1,1){\circle{4}} \put(0,1){$v_0$}
\put(16,1){\circle{4}} \put(15,1){$v_1$}
\put(31,1){\circle{4}} \put(30,1){$v_3$}
\put(46,1){\circle{4}} \put(45,1){$v_6$}

\put(3,1){\vector(1,0){10}} \put(7,-1){$\pi_1$}
\put(18,1){\vector(1,0){10}} \put(21,-1){$\pi_3$}
\put(33,1){\vector(1,0){10}} \put(38,-1){$\pi_6$}

\put(1,15){\circle{4}} \put(0,15){$v_2$}
\put(16,15){\circle{4}} \put(15,15){$v_5$}
\put(22,15){\circle{4}} \put(21,15){$v_4$}
\put(31,15){\circle{4}} \put(30,15){$v_7$}

\put(1,3){\vector(0,1){10}} \put(-4,7){$1-\pi_1$}
\put(16,3){\vector(0,1){10}} \put(7,7){$1-\pi_3-\pi_4$}
\put(16,3){\vector(1,2){5}} \put(19,6){$\pi_4$}
\put(31,3){\vector(0,1){10}} \put(32,7){$1-\pi_6$}
\end{picture}
\caption{Primitive probabilities and atomic event probabilities}
\label{simpletreeFig}
\end{figure}


\begin{figure}
\setlength{\unitlength}{0.4cm}
\begin{picture}(1,16)(9,-3)\thicklines
\put(0,5){\circle{1.5}} \put(0,5){\makebox(0,0){$v_0$}}

\put(5,9){\circle{1.5}} \put(5,9){\makebox(0,0){$v_1$}}
                        \put(5,10.3){\makebox(0,0){$A$}}
\put(5,1){\circle{1.5}} \put(5,1){\makebox(0,0){$v_2$}}
                        \put(5,2.3){\makebox(0,0){$\overline A$}}

\put(10,11){\circle{1.5}} \put(10,11){\makebox(0,0){$v_3$}}
                  \put(12.1,11){\makebox(0,0){$A\rightarrow B$}}
\put(10,9){\circle{1.5}} \put(10,9){\makebox(0,0){$v_4$}}
                  \put(12.1,9){\makebox(0,0){$A\rightarrow C$}}
\put(10,7){\circle{1.5}} \put(10,7){\makebox(0,0){$v_5$}}
                  \put(12.8,7){\makebox(0,0){$A\rightarrow\overline{B\cup C}$}}

\put(10,4){\circle{1.5}} \put(10,4){\makebox(0,0){$v_6$}}
                  \put(12.1,4){\makebox(0,0){$\overline A\rightarrow B$}}
\put(10,1){\circle{1.5}} \put(10,1){\makebox(0,0){$v_7$}}
                  \put(11.5,2.1){\makebox(0,0){$\overline A\rightarrow C$}}
\put(10,-2){\circle{1.5}} \put(10,-2){\makebox(0,0){$v_8$}}
               \put(12.8,-2){\makebox(0,0){$\overline
                   A\rightarrow\overline{B\cup C}$}}

\put(15,2){\circle{1.5}} \put(15,2){\makebox(0,0){$v_9$}}
                  \put(18.1,2){\makebox(0,0){$\overline{A}\rightarrow
                      C\rightarrow B$}}
\put(15,0){\circle{1.5}} \put(15,0){\makebox(0,0){$v_{10}$}}
             \put(18.1,0){\makebox(0,0){$\overline{A}\rightarrow
                 C\rightarrow\overline{B}$}}

\put(0.5,5){  \vector(1,-1){3.5} }
\put(0.5,5){  \vector(1,1){3.5} }

\put(5.5,9){  \vector(2,-1){3.2} }
\put(5.5,9){  \vector(1,0){3.2} }
\put(5.5,9){  \vector(2,1){3.2} }

\put(5.5,1){  \vector(4,-3){3.4} }
\put(5.5,1){  \vector(1,0){3.2} }
\put(5.5,1){  \vector(4,3){3.4} }

\put(10.5,1){  \vector(3,-1){3.2} }
\put(10.5,1){  \vector(3,1){3.2} }
\end{picture} \label{Fig1}
\caption{Stages and independence.}
\label{stages_indep}
\end{figure}
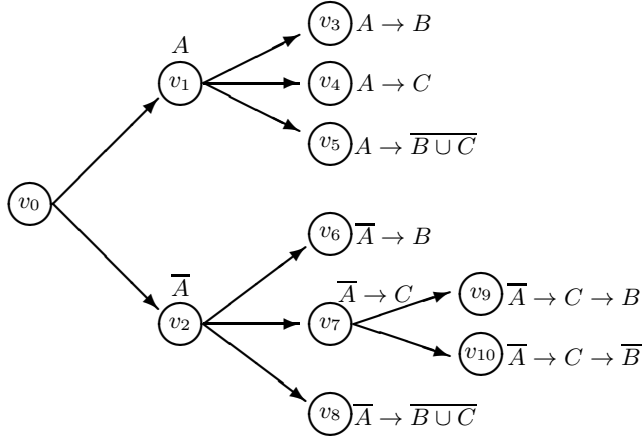


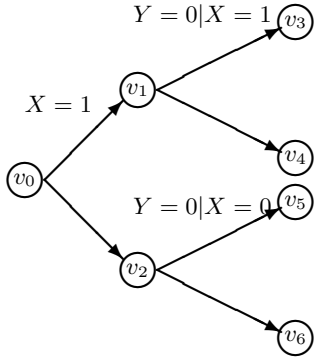
\begin{figure}
  \setlength{\unitlength}{0.3cm}
  \begin{picture}(1,18)(6,-3)\thicklines \put(0,5){\circle{1.5}}
    \put(0,5){\makebox(0,0){$v_0$}}

    \put(5,9){\circle{1.5}} \put(5,9){\makebox(0,0){$v_1$}}
    \put(5,1){\circle{1.5}} \put(5,1){\makebox(0,0){$v_2$}}

    \put(12,12){\circle{1.5}} \put(12,12){\makebox(0,0){$v_3$}}
    \put(12,6){\circle{1.5}} \put(12,6){\makebox(0,0){$v_4$}}

    \put(12,4){\circle{1.5}} \put(12,4){\makebox(0,0){$v_5$}}
    \put(12,-2){\circle{1.5}} \put(12,-2){\makebox(0,0){$v_6$}}

    \put(0.5,5){  \vector(1,-1){3.5} }\put(0,8){$X=1$}
    \put(0.5,5){  \vector(1,1){3.5} }

    \put(5.5,9){  \vector(2,-1){5.5}}\put(4.8,12){$Y=0|X=1$}
    \put(5.5,9){  \vector(2,1){5.5} }

    \put(5.5,1){  \vector(2,-1){5.5}}\put(4.8,3.5){$Y=0|X=0$}
    \put(5.5,1){  \vector(2,1){5.5} }

  \end{picture}
  \caption{Probability tree for Example \ref{simpledag}}
  \label{Figure1}
\end{figure}


\begin{figure}
  \setlength{\unitlength}{0.4cm}
  \begin{center}
    \begin{picture}(20,5)(0,4)\thicklines \put(0,5){\circle{2.5}}
      \put(0,5){\makebox(0,0){$[v_0]$}} \put(10,5){\circle{2.5}}
      \put(10,5){\makebox(0,0){$[v_1,v_2]$}} \put(20,5){\circle{2.5}}
      \put(20,5){\makebox(0,0){$w_\infty$}}
      \qbezier(1.3,5)(5,7)(8.4,5.2) \put(8.4,5.2){\vector(1,-1){0.2}}
      \qbezier(1.3,5)(5,3)(8.4,4.8) \put(8.4,4.8){\vector(1,1){0.2}}

      \qbezier(11.3,5)(15,7)(18.4,5.2) \put(18.4,5.2){\vector(1,-1){0.2}}
      \qbezier(11.3,5)(15,3)(18.4,4.8) \put(18.4,4.8){\vector(1,1){0.2}}
      \put(4,6.5){\text{$\pi(v_1)$}}
      \put(14,6.5){\text{$\pi(v_3)$}}
      \put(15.5,3){\text{$\pi(v_5)$}}
    \end{picture}
    \caption{CEG for Figure \ref{Figure1}}
    \label{CEGFigure1}
  \end{center}
\end{figure}
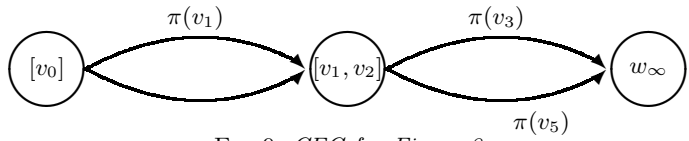

\begin{figure}
\setlength{\unitlength}{0.3cm}
\begin{picture}(4,32)(9,-21)\thicklines
\put(9,10){\circle{1.5}} \put(9,10){\makebox(0,0){$v_0$}}

\put(8.7,9.2){  \vector(-2,-1){7.8} }
\put(8.7,9.2){  \vector(0,-1){4} }
\put(8.7,9.2){  \vector(2,-1){7.8} }
\put(1,4.5){\circle{1.5}} \put(1,4.5){\makebox(0,0){$v_1$}}
\put(9,4.5){\circle{1.5}} \put(9,4.5){\makebox(0,0){$v_2$}}
\put(17,4.5){\circle{1.5}} \put(17,4.5){\makebox(0,0){$v_3$}}

\put(0.7,3.7){  \vector(-1,-2){2} }
\put(0.7,3.7){  \vector(1,-2){2} }

\put(8.7,3.7){  \vector(-1,-2){2} }
\put(8.7,3.7){  \vector(1,-2){2} }

\put(16.7,3.7){  \vector(-1,-2){2} }
\put(16.7,3.7){  \vector(1,-2){2} }

\put(-1.1,-1){\circle{1.5}} \put(-1.1,-1){\makebox(0,0){$v_4$}}
\put(3.1,-1){\circle{1.5}} \put(3.1,-1) {\makebox(0,0){$v_5$}}

\put(7,-1){\circle{1.5}} \put(7,-1){\makebox(0,0){$v_6$}}
\put(11,-1){\circle{1.5}} \put(11,-1) {\makebox(0,0){$v_7$}}

\put(15,-1){\circle{1.5}} \put(15,-1){\makebox(0,0){$v_8$}}
\put(19,-1){\circle{1.5}} \put(19,-1) {\makebox(0,0){$v_9$}}

\put(2.9,-1.8){  \vector(-1,-2){2} }
\put(2.9,-1.8){  \vector(1,-2){2} }

\put(10.9,-1.8){  \vector(-1,-2){2} }
\put(10.9,-1.8){  \vector(1,-2){2} }

\put(18.9,-1.8){  \vector(-1,-2){2} }
\put(18.9,-1.8){  \vector(1,-2){2} }

\put(1.2,-6.5){\circle{1.5}} \put(1.2,-6.5){\makebox(0,0){$v_{10}$}}
\put(5.2,-6.5){\circle{1.5}} \put(5.2,-6.5) {\makebox(0,0){$v_{11}$}}

\put(9.2,-6.5){\circle{1.5}} \put(9.2,-6.5){\makebox(0,0){$v_{12}$}}
\put(13.2,-6.5){\circle{1.5}} \put(13.2,-6.5){\makebox(0,0){$v_{13}$}}

\put(17.2,-6.5){\circle{1.5}} \put(17.2,-6.5){\makebox(0,0){$v_{14}$}}
\put(21.2,-6.5){\circle{1.5}} \put(21.2,-6.5) {\makebox(0,0){$v_{15}$}}

\put(12.9,-7.2){  \vector(-1,-2){2} }
\put(12.9,-7.2){  \vector(1,-2){2} }

\put(11.1,-12){\circle{1.5}} \put(11.1,-12){\makebox(0,0){$v_{16}$}}
\put(15.3,-12){\circle{1.5}} \put(15.3,-12) {\makebox(0,0){$v_{17}$}}

\put(15,-12.8){  \vector(-1,-2){2} }
\put(15,-12.8){  \vector(1,-2){2} }

\put(13.1,-17.5){\circle{1.5}} \put(13.1,-17.5){\makebox(0,0){$v_{18}$}}
\put(17.3,-17.5){\circle{1.5}} \put(17.3,-17.5) {\makebox(0,0){$v_{19}$}}

\put(17,-18.3){  \vector(-2,-1){3.8} }
\put(17,-18.3){  \vector(0,-1){2} }
\put(17,-18.3){  \vector(2,-1){3.8} }
\put(13.3,-21){\circle{1.5}} \put(13.3,-21){\makebox(0,0){$v_{20}$}}
\put(17.3,-21){\circle{1.5}} \put(17.3,-21) {\makebox(0,0){$v_{21}$}}
\put(21.3,-21){\circle{1.5}} \put(21.3,-21){\makebox(0,0){$v_{22}$}}

\end{picture}
\caption{A tree with unusual stage set}
\label{fax.eps}
\end{figure}
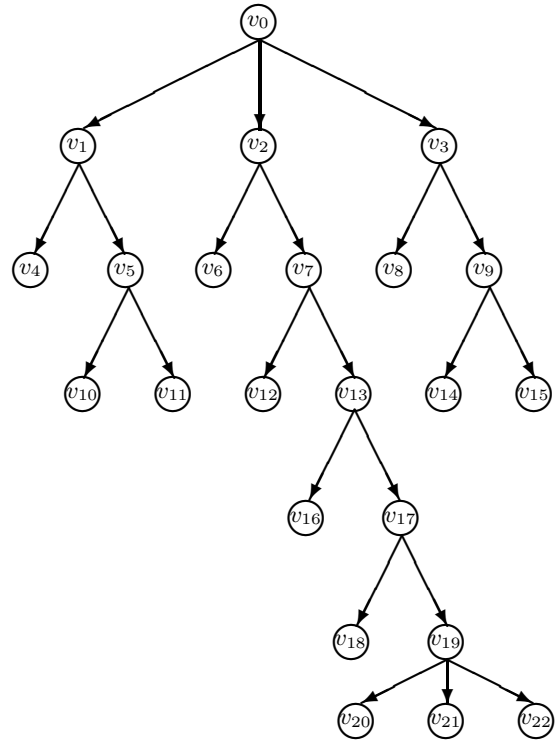


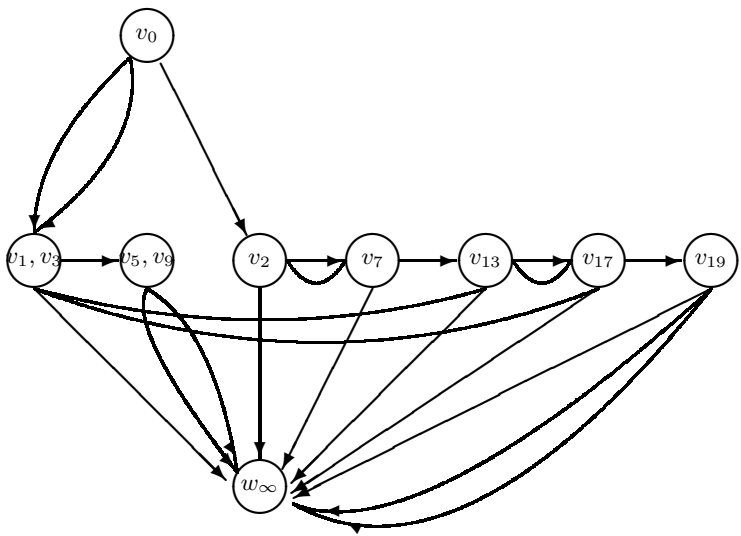
\begin{figure}
\setlength{\unitlength}{0.3cm}
\begin{picture}(50,25)(2,-3)\thicklines
\put(15,0){\circle{2.5}}  \put(15,0){\makebox(0,0){$w_{\infty}$}}
\put(5,10){\circle{2.5}}  \put(5,10){\makebox(0,0){$v_1,v_3$}}
\put(10,10){\circle{2.5}}  \put(10,10){\makebox(0,0){$v_5,v_9$}}
\put(15,10){\circle{2.5}}  \put(15,10){\makebox(0,0){$v_2$}}
\put(20,10){\circle{2.5}}  \put(20,10){\makebox(0,0){$v_7$}}
\put(25,10){\circle{2.5}}  \put(25,10){\makebox(0,0){$v_{13}$}}
\put(30,10){\circle{2.5}}  \put(30,10){\makebox(0,0){$v_{17}$}}
\put(35,10){\circle{2.5}}  \put(35,10){\makebox(0,0){$v_{19}$}}
\put(10,20){\circle{2.5}}  \put(10,20){\makebox(0,0){$v_0$}}

\put(10.25,18.75){  \vector(1,-2){3.8} }
\qbezier(9.25,19)(5,15)(5,11.25) \put(5,11.5){\vector(0,-2){0.2}}
\qbezier(9.25,19)(10,15)(5,11.25) \put(5.4,11.5){\vector(-2,-1){0.2}}

\put(6.2,10){\vector(1,0){2.5}}
\put(16.2,10){\vector(1,0){2.5}}
\put(21.2,10){\vector(1,0){2.5}}
\put(26.2,10){\vector(1,0){2.5}}
\put(31.2,10){\vector(1,0){2.5}}

\qbezier(16.2,10)(17.5,8)(18.8,10)
\qbezier(26.2,10)(27.5,8)(28.8,10)

\qbezier(5,8.75)(15.4,6)(25,8.75)
\qbezier(5,8.75)(17.9,4)(30,8.75)

\put(5,8.75){\vector(1,-1){8.5}}
\put(15,8.75){\vector(0,-1){7.5}}
\put(20,8.75){\vector(-1,-2){4}}
\put(25,8.75){\vector(-1,-1){8.6}}
\put(30,8.75){\vector(-3,-2){13.6}}

\qbezier(10,8.75)(9,7)(14,0.6)  \put(13.7,1.0){\vector(1,-1){0.2}}
\qbezier(10,8.75)(13,7)(14,0.6) \put(13.8,1.6){\vector(2,-3){0.2}}

\put(35,8.75){\vector(-2,-1){18.5}}
\qbezier(35,8.75)(22,-3)(16.5,-0.8) \put(18,-1.1){\vector(-1,0){0.2}}
\qbezier(35,8.75)(24,-5)(16.5,-0.8) \put(19,-1.7){\vector(-2,1){0.2}}

\end{picture}
\caption{CEG for the event tree in Figure \ref{fax.eps}}
\label{CEGfax.eps}
\end{figure}


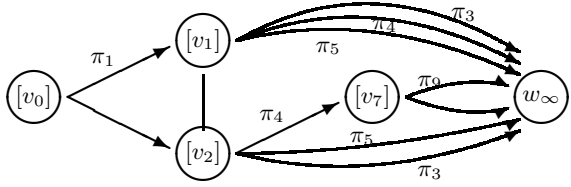
\begin{figure}
\setlength{\unitlength}{0.15cm}
\begin{picture}(50,17)(-1,-1)\thicklines

\put(0,5){\circle{5}} \put(0,5){\makebox(0,0){$[v_0]$}}

\put(15,0){\circle{5}} \put(15,0){\makebox(0,0){$[v_2]$}}
\put(15,10){\circle{5}} \put(15,10){\makebox(0,0){$[v_1]$}}

\put(3,5){\vector(2,-1){9}}
\put(3,5){\vector(2,1){9}} \put(5,8){$\pi_1$}

\put(15,2){\line(0,1){5}}  \put(20,3){$\pi_4$}
\put(30,5){\circle{5}} \put(30,5){\makebox(0,0){$[v_7]$}}
\put(18,0){\vector(2,1){9}}

\qbezier(33,5)(38,7)(42,6)  \put(42,6){\vector(2,-1){0.2}}
   \put(34,6){$\pi_9$}
\qbezier(33,5)(38,3)(42,4)  \put(42,4){\vector(2,1){0.2}}

\put(45,5){\circle{5}} \put(45,5){\makebox(0,0){$w_\infty$}}

\qbezier(18,0)(30,0)(43,3)  \put(43,3){\vector(2,1){0.2}}
   \put(34,-2){$\pi_3$}
\qbezier(18,0)(30,-3)(43,2)  \put(43,2){\vector(2,1){0.2}}
   \put(28,1){$\pi_5$}

\qbezier(18,10)(30,17)(43,9)  \put(43,9){\vector(2,-1){0.2}}
   \put(37,12){$\pi_3$}
\qbezier(18,10)(30,15)(43,8)  \put(43,8){\vector(2,-1){0.2}}
   \put(30,11){$\pi_4$}
\qbezier(18,10)(30,13)(43,7)  \put(43,7){\vector(2,-1){0.2}}
   \put(25,9){$\pi_5$}

\end{picture}
\caption{CEG for the event tree in Figure \ref{stages_indep}}
\label{ABCceg}
\end{figure}


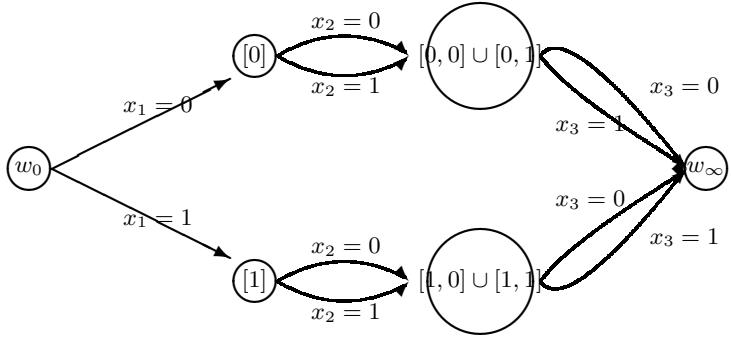
\begin{figure}
\setlength{\unitlength}{0.25cm}
\begin{picture}(30,15)(2,-9)\thicklines

\put(0,0){\circle{2.5}} \put(0,0){\makebox(0,0){$w_{0}$}}
\put(12,6){\circle{2.5}} \put(12,6){\makebox(0,0){$[0]$}}
\put(12,-6){\circle{2.5}} \put(12,-6){\makebox(0,0){$[1]$}}
\put(24,6){\circle{8.5}} \put(24,6){\makebox(0,0){$[0,0]\cup[0,1]$}}
\put(24,-6){\circle{8.5}} \put(24,-6){\makebox(0,0){$[1,0]\cup[1,1]$}}
\put(36,0){\circle{2.5}} \put(36,0){\makebox(0,0){$w_\infty$}}

\put(1.2,0){\vector(2,1){9.5}}   \put(5,3){$x_1=0$}
\put(1.2,0){\vector(2,-1){9.5}}  \put(5,-3){$x_1=1$}

\qbezier(13.2,6)(16.6,8)(20,6.2)    \put(20,6.2){\vector(1,-1){0.2}}  \put(15,7.5){$x_2=0$}
\qbezier(13.2,6)(16.6,4)(20,5.8)    \put(20,5.8){\vector(1,1){0.2}}   \put(15,4){$x_2=1$}
\qbezier(13.2,-6)(16.6,-4)(20,-5.8) \put(20,-5.8){\vector(1,-1){0.2}} \put(15,-4.5){$x_2=0$}
\qbezier(13.2,-6)(16.6,-8)(20,-6.2) \put(20,-6.2){\vector(1,1){0.2}}  \put(15,-8){$x_2=1$}

\qbezier(27.2,6)(28.6,8)(34.8,0.2)    \put(34.8,0.2){\vector(1,-1){0.2}}  \put(33,4){$x_3=0$}
\qbezier(27.2,6)(28.6,4)(34.8,0.2)    \put(34.8,0.2){\vector(1,1){0.2}}   \put(28,2){$x_3=1$}
\qbezier(27.2,-6)(28.6,-4)(34.8,-0.2) \put(34.8,-0.2){\vector(1,-1){0.2}} \put(28,-2){$x_3=0$}
\qbezier(27.2,-6)(28.6,-8)(34.8,-0.2) \put(34.8,-0.2){\vector(1,1){0.2}}  \put(33,-4){$x_3=1$}

\end{picture}
\caption{CEG for binary  $X_{2}\leftarrow X_{1}\rightarrow X_{3}$}
\label{fromCEGtoBNfigure}
\end{figure}


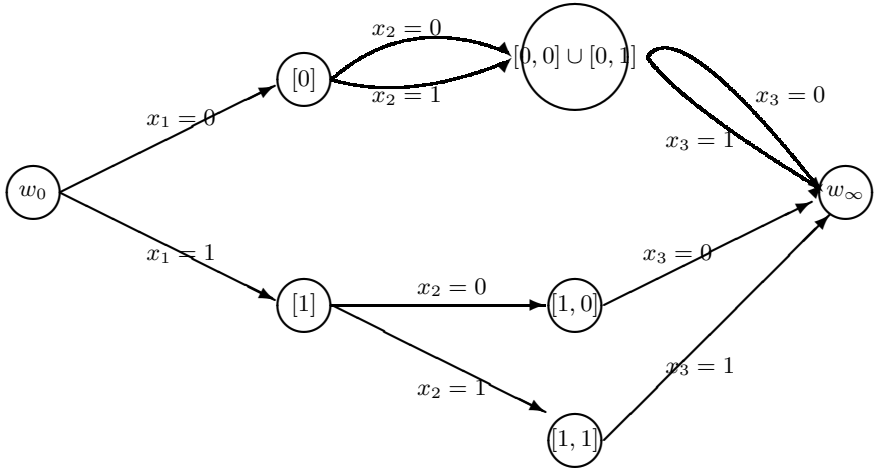
\begin{figure}
\setlength{\unitlength}{0.3cm}
\begin{picture}(50,23)(2,-13)\thicklines

\put(0,0){\circle{2.5}} \put(0,0){\makebox(0,0){$w_{0}$}}
\put(12,5){\circle{2.5}} \put(12,5){\makebox(0,0){$[0]$}}
\put(12,-5){\circle{2.5}} \put(12,-5){\makebox(0,0){$[1]$}}
\put(24,6){\circle{8.5}} \put(24,6){\makebox(0,0){$[0,0]\cup[0,1]$}}
\put(24,-5){\circle{2.5}} \put(24,-5){\makebox(0,0){$[1,0]$}}
\put(24,-11){\circle{2.5}} \put(24,-11){\makebox(0,0){$[1,1]$}}
\put(36,0){\circle{2.5}} \put(36,0){\makebox(0,0){$w_\infty$}}

\put(1.2,0){\vector(2,1){9.5}}   \put(5,3){$x_1=0$}
\put(1.2,0){\vector(2,-1){9.5}}  \put(5,-3){$x_1=1$}

\qbezier(13.2,5)(16.6,8)(21,6.2) \put(21,6.2){\vector(1,-1){0.2}} \put(15,7){$x_2=0$}
\qbezier(13.2,5)(16.6,4)(21,5.8) \put(21,5.8){\vector(1,1){0.2}}  \put(15,4){$x_2=1$}
\put(13.2,-5){\vector(1,0){9.5}}  \put(17,-4.5){$x_2=0$}
\put(13.2,-5){\vector(2,-1){9.5}} \put(17,-9){$x_2=1$}

\qbezier(27.2,6)(28.6,8)(34.8,0.2)    \put(34.8,0.2){\vector(1,-1){0.2}}  \put(32,4){$x_3=0$}
\qbezier(27.2,6)(28.6,4)(34.8,0.2)    \put(34.8,0.2){\vector(1,1){0.2}}   \put(28,2){$x_3=1$}
\put(25.3,-5){\vector(2,1){9.2}}    \put(27,-3){$x_3=0$}
\put(25.3,-11){\vector(1,1){10}}  \put(28,-8){$x_3=1$}

\end{picture}
\caption{CEG for condition (\ref{conditone}) but not necessarily (\ref{condit2})}
\label{fromCEGtoBNmodification}
\end{figure}


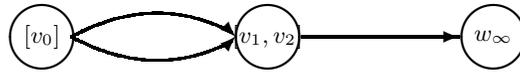
\begin{figure}
\setlength{\unitlength}{0.3cm}
\begin{center}
\begin{picture}(20,5)(0,4)\thicklines
\put(0,5){\circle{2.8}} \put(0,5){\makebox(0,0){$[v_0]$}}
\put(10,5){\circle{2.8}} \put(10,5){\makebox(0,0){$[v_1,v_2]$}}
\put(20,5){\circle{2.8}} \put(20,5){\makebox(0,0){$w_\infty$}}
\qbezier(1.3,5)(5,7)(8.4,5.2) \put(8.4,5.2){\vector(1,-1){0.2}}
\qbezier(1.3,5)(5,3)(8.4,4.8) \put(8.4,4.8){\vector(1,1){0.2}}

\put(11.5,5){\vector(1,0){7}}

\end{picture}
\caption{Manipulated CEG for Example \ref{Examplesimpledagmanipulated}}
\label{simpledagmanipulated}
\end{center}
\end{figure}


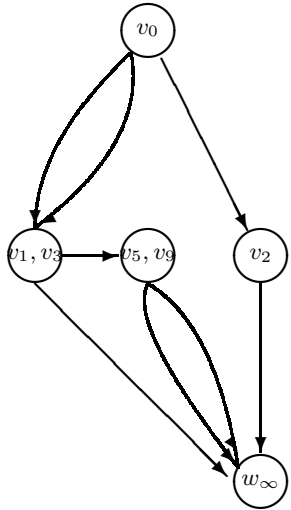
\begin{figure}
\setlength{\unitlength}{0.3cm}
\begin{picture}(30,23)(-2,-2)\thicklines
\put(15,0){\circle{2.5}}  \put(15,0){\makebox(0,0){$w_{\infty}$}}
\put(5,10){\circle{2.5}}  \put(5,10){\makebox(0,0){$v_1,v_3$}}
\put(10,10){\circle{2.5}}  \put(10,10){\makebox(0,0){$v_5,v_9$}}
\put(15,10){\circle{2.5}}  \put(15,10){\makebox(0,0){$v_2$}}

\put(10,20){\circle{2.5}}  \put(10,20){\makebox(0,0){$v_0$}}

\put(10.25,18.75){  \vector(1,-2){3.8} }
\qbezier(9.25,19)(5,15)(5,11.25) \put(5,11.5){\vector(0,-2){0.2}}
\qbezier(9.25,19)(10,15)(5,11.25) \put(5.4,11.5){\vector(-2,-1){0.2}}

\put(6.2,10){\vector(1,0){2.5}}

\put(5,8.75){\vector(1,-1){8.5}}
\put(15,8.75){\vector(0,-1){7.5}}

\qbezier(10,8.75)(9,7)(14,0.6)  \put(13.7,1.0){\vector(1,-1){0.2}}
\qbezier(10,8.75)(13,7)(14,0.6) \put(13.8,1.6){\vector(2,-3){0.2}}

\end{picture}
\caption{Manipulated CEG for Example \ref{ManipulatedFAXbis}}
\label{ManipulatedFAX.Picture}
\end{figure}


\begin{figure}[htbp]
\begin{center}
\epsfig{height=10truecm,width=7truecm,file=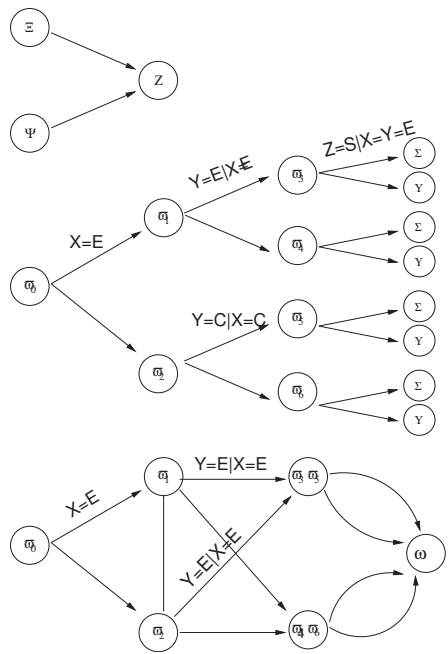}
\end{center}
\caption{BN and CEG for the example of university room allocation}
\label{University.eps}
\end{figure}


\begin{figure}[htbp]
\begin{center}
\epsfig{height=10truecm,width=12truecm,file=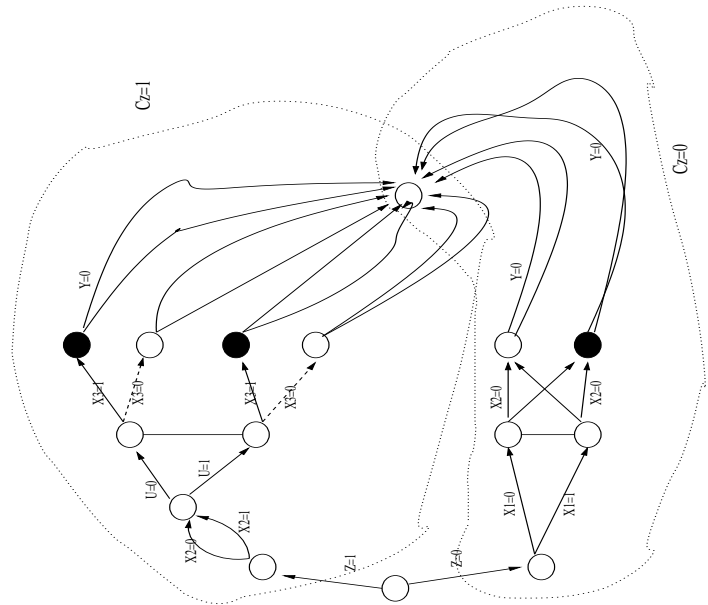}
\end{center}
\caption{Modified university example}
\label{CEGsimpleBDTaaa}
\end{figure}

\end{document}